\documentclass[twocolumn]{aastex701}

\newcommand{\phsup}{\phantom{$^{0}$}}
\newcommand{\fdeg}{.\!^{\circ}}

\usepackage{textcomp}
\usepackage{subcaption}
\usepackage[flushleft]{threeparttable}
\usepackage{amsmath}
\usepackage[T1]{fontenc}

\submitjournal{ApJ Letters}

\begin{document}

\title{A quest for sulfur-bearing refractory species. \\* Identification of CaS in the interstellar medium}

\author[orcid=0009-0002-3398-4627]{A.\ Tasa-Chaveli}
\affiliation{Centro de Astrobiolog\'ia (CAB), CSIC-INTA, Ctra.\ de Ajalvir, km 4, Torrej\'on de Ardoz, 28850 Madrid, Spain}
\email[show]{atasa@cab.inta-csic.es}

\author[orcid=0000-0002-3078-9482]{\'A.\ S\'anchez-Monge}
\affiliation{Institut de Ci\`encies de l’Espai (ICE), CSIC, Campus UAB, Carrer de Can Magrans s/n, 08193 Bellaterra, Barcelona, Spain}
\affiliation{Institut d'Estudis Espacials de Catalunya (IEEC), 08860 Castelldefels, Barcelona, Spain}
\email[show]{asanchez@ice.csic.es}

\author[orcid=0000-0001-6317-6343]{A.\ Fuente}
\affiliation{Centro de Astrobiolog\'ia (CAB), CSIC-INTA, Ctra.\ de Ajalvir, km 4, Torrej\'on de Ardoz, 28850 Madrid, Spain}
\email{afuente@cab.inta-csic.es}

\author[orcid=0000-0001-6431-9633]{A.\ Ginsburg}
\affiliation{Department of Astronomy, University of Florida, P.O.\ Box 112055, Gainesville, FL 32611, USA}
\email{adam.g.ginsburg@gmail.com}

\author[orcid=0000-0002-0183-8927]{H.~S.~P.\ M\"uller}
\affiliation{I.\ Physikalisches Institut, Universit\"at zu K\"oln, Z\"ulpicher Str.\ 77, 50937 K\"oln, Germany}
\email{hspm@ph1.uni-koeln.de}

\author[orcid=0000-0002-9277-8025]{Th.\ M\"oller}
\affiliation{I.\ Physikalisches Institut, Universit\"at zu K\"oln, Z\"ulpicher Str.\ 77, 50937 K\"oln, Germany}
\email{moeller@ph1.uni-koeln.de}

\author[orcid=0000-0003-0969-8137]{P.\ Rivi\`ere-Marichalar}
\affiliation{Observatorio Astron\'omico Nacional (OAN), C.\ Alfonso XII, 3, 28014 Madrid, Spain}
\email{p.riviere@oan.es}

\author[orcid=0000-0002-8499-7447]{D.\ Navarro-Almaida}
\affiliation{Centro de Astrobiolog\'ia (CAB), CSIC-INTA, Ctra.\ de Ajalvir, km 4, Torrej\'on de Ardoz, 28850 Madrid, Spain}
\email{dnavarro@cab.inta-csic.es}

\author[orcid=0000-0002-4292-4127]{G.\ Esplugues}
\affiliation{Observatorio Astron\'omico Nacional (OAN), C.\ Alfonso XII, 3, 28014 Madrid, Spain}
\email{g.esplugues@oan.es}

\author[orcid=0009-0002-3398-4627]{P.\ Schilke}
\affiliation{I.\ Physikalisches Institut, Universit\"at zu K\"oln, Z\"ulpicher Str.\ 77, 50937 K\"oln, Germany}
\email{schilke@ph1.uni-koeln.de}

\author[orcid=0000-0002-7501-0073]{M.\ Rodr\'iguez-Baras}
\affiliation{European Space Agency (ESA), European Space Astronomy Centre (ESAC), Camino Bajo del Castillo s/n, 28692 Villanueva de la Cañada, Madrid, Spain}
\email{marina.rodriguezbaras@esa.int}

\author[orcid=0000-0001-8200-6710]{S.\ Thorwirth}
\affiliation{I.\ Physikalisches Institut, Universit\"at zu K\"oln, Z\"ulpicher Str.\ 77, 50937 K\"oln, Germany}
\email{sthorwirth@ph1.uni-koeln.de}

\author[orcid=0000-0003-0833-4075]{L.\ Beitia-Antero}
\affiliation{Departamento de Estad\'istica e Investigaci\'on Operativa, Facultad de Ciencias Matem\'aticas, Universidad Complutense de Madrid, Spain}
\affiliation{Joint Center for Ultraviolet Astronomy, Universidad Complutense de Madrid, Avda.\ Puerta de Hierro s/n, 28040 Madrid, Spain}
\email{lbeitia@ucm.es}

\begin{abstract}

The recent detection of refractory molecules in massive star-forming regions provides a means of probing the innermost regions of disks around massive stars. These detections also make it possible to explore the chemical composition of refractories through gas-phase observations. In this regard, identifying refractory compounds containing sulfur could reveal potential connections between sulfur and refractories, as well as help determine the sulfur budget in these extreme environments. We find convincing evidence of a reliable detection of CaS, and tentative detections of KS and KSH in the disk G351.77-mm1. These are the first ever identifications of these species in the interstellar medium. The CaS, KS, and KSH column densities are about 3 orders of magnitude lower than those of the abundant sulfur compounds  SO$_2$, CH$_3$SH and SiS, proving that these species are not the major reservoir of sulfur at the spatial scales probed by our observations. Higher angular resolution observations at different wavelengths are required to confirm these detections, which are of paramount importance to gain insights into the formation of gas-phase refractory molecules.

\end{abstract}


%
\section{Introduction}\label{sec:introduction}

Refractory species are compounds characterized by high sublimation temperatures ($\gtrsim700$~K). They are typically carbonaceous grains, silicates, and compounds with heavy elements (e.g., Na, K, Ca, Mg, Al, Fe) that are believed to be in a solid state in most phases of the interstellar medium \citep[see e.g.,][]{Das2025}. These conditions make them difficult to detect because the extreme conditions required for them to transition from the solid-phase to the gas-phase are often not met (e.g., extremely high temperatures, intense radiation, strong shocks). This has limited their search and study in the past to the envelopes of AGB stars \citep[e.g.,][]{Cernicharo2019, Wallstrom2024}. Interestingly, the recent detection of the refractory molecules NaCl and KCl in the protostellar disk of Orion~Src\,I \citep{Ginsburg2019} opened the door to search for other refractory compounds in star-forming regions. Since the first detection of NaCl and KCl in Orion~Src\,I, a few other disks have been found to emit in NaCl, KCl and AlO \citep[e.g.,][]{Tanaka2020, Ginsburg2023}. Along this line, sulfur-bearing refractory compounds, such as MgS and NaS, have been recently detected in the G+0.693$-$0.027 molecular cloud, suggesting a close relationship between sulfur and refractory elements \citep{ReyMontejo2024}.

Detecting and studying refractory molecules provides the possibility to probe regions and conditions that have remained hidden until now. In particular, they are known to trace the innermost regions of massive protostellar sources, including their inner accretion disks and material affected by powerful jets \citep[see][]{Tanaka2020}, enabling the study of the physical properties and dynamics down to $\sim100$~au. From a chemical perspective, refractory compounds are linked to the study of meteorites. Calcium- and aluminum-rich inclusions (CAIs) are considered the oldest materials in the solar system, but their formation processes are not yet understood \citep{Woitke2024, Jongejan2023}. Thus, studying whether refractory species may sublimate in star-forming regions can provide insights into the formation of these high-temperature meteoritic components. Furthermore, the detection of refractory species containing sulfur may also help to answer the long-standing question in astrochemistry about the ``missing sulfur problem'' in the dense gas interstellar phase.

This problem deals with sulfur, one of the most abundant elements in the Universe \citep[S/H $\sim1.5\times10^{-5}$,][]{Asplund2009} and known to play a significant role in biological systems \citep{Ranjan2022}. Despite its relevance, the sulfur chemistry has posed difficulties for decades. A significant challenge is that, although the gaseous sulfur accounts for its total cosmic abundance in diffuse clouds \citep[e.g.,][]{Neufeld2015} and photodissociation regions \citep[e.g.,][]{GoicoecheaCuadrado2021, Fuente2024, Fuente2025}, there is an unexpected shortage of sulfur within dense molecular clouds. In dense cores, the sum of the observed gas-phase abundances of S-bearing molecules constitutes only $<1\%$ of the expected amount \citep[e.g.,][]{Ruffle1999, Wakelam2004}. One could think that most of the sulfur must be locked on the icy mantles that cover dust grains. However, a similar trend is encountered there, where the observed abundances in the ice solid phase (s-OCS: \citealp{Palumbo1995}; s-SO$_2$: \citealp{Boogert1997}; s-H$_2$S upper limit: \citealp{JimenezEscobar2011}) account for only $<5\%$ of the total amount. Recent observations combined with theoretical and laboratory modeling have revised these numbers through a wide sample of starless dense cores \citep{Fuente2019, Fuente2023, Esplugues2025}. The new results confirm a high depletion of S-bearing species, setting a depletion factor of $\approx10$ for the studied regions, which is similar to that measured in hot cores and comets \citep[e.g.,][]{Calmonte2016}. Based on these findings, it is hypothesized that the so-called depleted sulfur, which adds up to $\approx90\%$ of the total sulfur, might be locked in refractory or (semi)refractory material (e.g., S-bearing species containing Na, K, Ca, Mg), as well as large sulfur allotropes \citep[e.g., S$_3$, S$_4$, S$_8$;][] {Shingledecker2020, Cazaux2022}. The search and detection of S-bearing refractory species may provide key information to understand the fate of sulfur in the dense gas interstellar medium.

This letter presents a search for sulfur compounds containing refractory elements in the disk around the massive young stellar object G351.77$-$0.54-mm1 (hereafter G351.77-mm1) based on high spatial resolution observations carried out with the Atacama Large Millimeter/sub-millimeter Array (ALMA).

\begin{table}
\begin{center}
\caption{Spectral setup of the ALMA observations\label{tab:spw-info}}
\begin{tabular}{c c c c}
\tableline\tableline \noalign{\smallskip}
  spw
& Freq.\ range
& Beam size; PA
& rms
\\
\tableline \noalign{\smallskip}
25  & 186.885--188.760 & $0.224\times0.167$; $-82.4$ & 1.52 \\
27  & 188.754--190.629 & $0.224\times0.161$; $-83.8$ & 1.19 \\
29  & 199.837--200.712 & $0.221\times0.151$; $-86.3$ & 0.96 \\
31  & 200.706--202.581 & $0.213\times0.149$; $-87.1$ & 0.96 \\
\tableline   
\end{tabular}
\end{center}
\tablecomments{The columns list the spectral window (spw) id number; the frequency range in GHz; the synthesized beam size in arcsec and its position angle in degrees; and the rms in mJy~beam$^{-1}$ (multiply by 0.93 to convert to K).}
\end{table}

%
\section{Observations}\label{sec:observations}

\begin{figure}
\centering
\includegraphics[width=1.0\columnwidth]{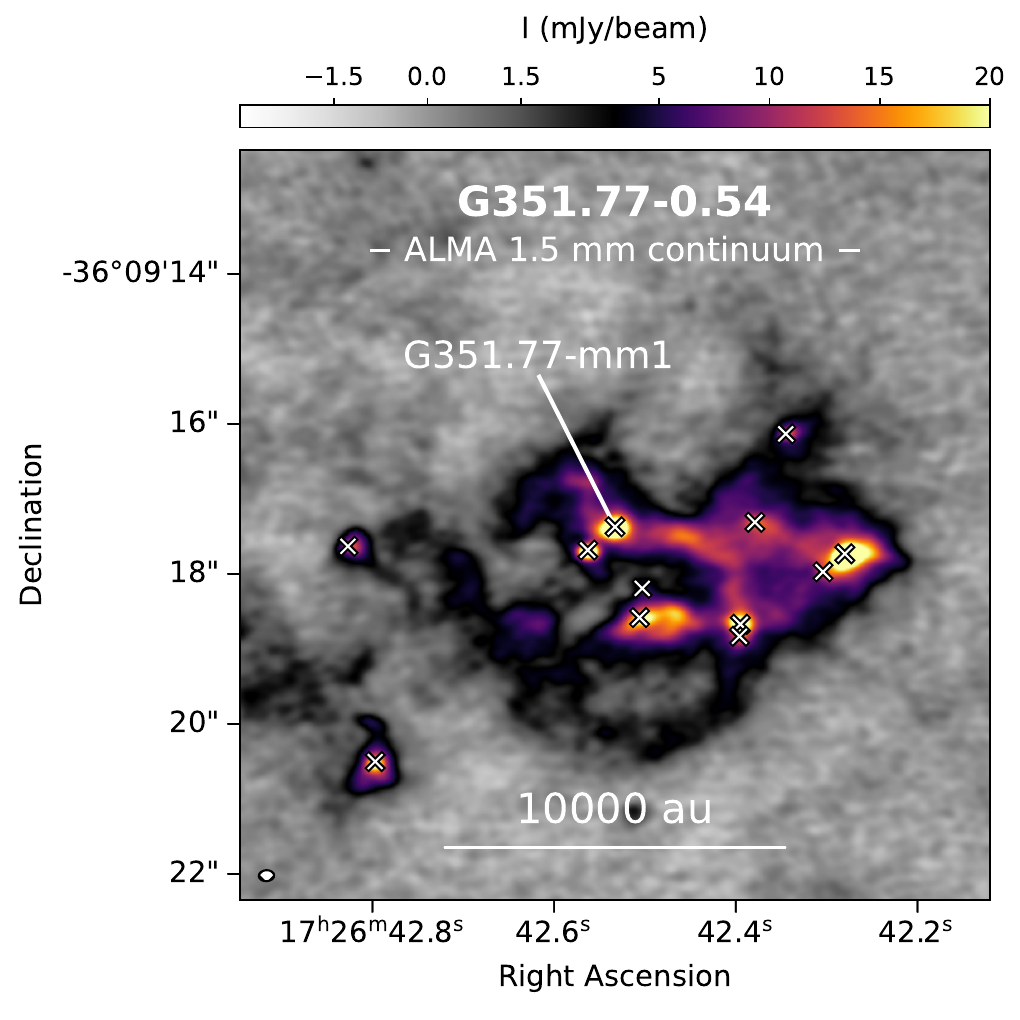}
\vspace{-0.5cm}
\caption{ALMA 1.5~mm continuum emission towards the G351.77$-$0.54 star-forming region. Crosses mark the positions of the mm continuum sources identified by \citet[][]{Beuther2019}. The synthesized beam of the image (beam $=0\farcs21\times0\farcs15$, PA$=-86^\circ$; with an $\mathrm{rms}\simeq0.32$~mJy~beam$^{-1}$) is shown in the bottom-left corner. The spectra shown in Figs.~\ref{fig:zooms} and \ref{fig:total-spectrum} have been extracted towards the peak position of G351.77-mm1, located at $\alpha(\mathrm{J2000})=17^\mathrm{h}26^\mathrm{m}42.533^\mathrm{s}$ and $\delta(\mathrm{J2000})=-36^{\circ}9^{\prime}17.37^{\prime\prime}$.}
\label{fig:continuum}
\end{figure}

The data presented in this work were observed with ALMA in December 2023 as part of a cycle 10 project 2023.1.01382.S (PI: \'A.\ S\'anchez-Monge), using 44 antennas with baselines between 15~m and 2.5~km providing an angular resolution of $\approx0\farcs16$. The observations were carried out in Band~5, covering parts of the frequency range from 186 to 203~GHz with four spectral windows (see Table~\ref{tab:spw-info}), each with a bandwidth of 1875~MHz, and a channel velocity resolution of $\sim$1.5~km~s$^{-1}$. The phase center of the G351.77-mm1 observations was set to $\alpha$(ICRS)=$17^{\rm h}26^{\rm m}42.\!^{\rm s}5330$ and $\delta$(ICRS)=$-36^{\circ}09^\prime17\farcs345$. Phase calibration was performed using the quasar J1720$-$3552, while flux and bandpass calibrations were obtained through observations of J1427$-$4206. The amount of precipitable water vapor was about 1.68~mm. Data were calibrated and imaged using {\sc casa} Version 6.5.4.9. We created cubes with the robust 0.5 weighting. The resulting synthesized beams and rms noise levels are listed in Table~\ref{tab:spw-info}. We also generate a map of the continuum emission at 1.5~mm using the line-free channels identified using the {\sc casa} funtion \texttt{findcont}. The resulting continuum image, showin in Fig.~\ref{fig:continuum}, has a rms noise level of 0.32~mJy~beam$^{-1}$, and a synthesized beam of $0\farcs206\times0\farcs148$, with a PA$=-86\fdeg2$.

%
\section{Fitting the spectra}\label{sec:results}

G351.77-mm1 is one of the brightest disks detected in NaCl by \citet{Ginsburg2023}. The size of the protostellar disk, of $\sim$$0\farcs108$ ($\sim$240~au with an adopted distance of 2.2~kpc; \citealp{Reid2014}) is similar to the Half Power Beam Width (HPBW) of our observations synthesized beam, thus optimizing the search for new species. Figure~\ref{fig:continuum} shows the ALMA 1.5~mm continuum emission and marks the positions of the dense cores reported by \citet{Beuther2019}.
The continuum-subtracted spectrum extracted towards the peak position of the disk G351.77-mm1 (with coordinates $\alpha(\mathrm{J2000})=17^\mathrm{h}26^\mathrm{m}42.533^\mathrm{s}$, $\delta(\mathrm{J2000})=-36^{\circ}9^{\prime}17.37^{\prime\prime}$) is shown in Fig.~\ref{fig:total-spectrum}. As a first step of our analysis, we have identified and fitted the lines of the major carriers in these spectral windows. Line identification was carried out by searching in the Cologne Database for Molecular Spectroscopy \citep[CDMS;][]{Mueller2005} and the Jet Propulsion Laboratory Line catalogues \citep[JPL;][]{Pickett1998} for intense transitions within our frequency range and generating synthetic spectra using the eXtended CASA Line Analysis Software Suite \citep[XCLASS;][]{Moeller2017}, which assumes LTE excitation conditions. We employed the Trust Region Reflective algorithm (available within XCLASS) to find the parameters that best describe the data. The five main parameters to be fitted with XCLASS are the source size, the molecular column density, the temperature, the velocity linewidth, and the velocity offset. For the source size, we used a fixed value of $0\farcs1$ based on the moment-0 maps of the resolved NaCl emission presented by \citet{Ginsburg2023}. The other parameters were fitted when possible or fixed to a certain value as described below. We fitted $\sim$90\% of the most intense lines ($T>$50~K) observed in the spectrum after including 21 species and their isotopologues (see Table~\ref{tab:other-species}). Figure~\ref{fig:total-spectrum} shows the best fit to the global spectrum of G351.77-mm1. When a significant number of lines were identified, the rotational temperature was fitted. In cases where this was not possible, we assumed $T_{\rm rot}=400$~K, as an average value of the temperatures derived for the species with multiple detected transitions (e.g., HC$_3$N: $486\substack{+14 \\ -147}$~K, CH$_3$CN: $300\substack{+5 \\ -5}$~K, C$_2$H$_3$CN: $378\substack{+23 \\ -7}$~K, C$_2$H$_5$CN: $473\substack{+32 \\ -82}$~K, CH$_3$OH: $473\substack{+21 \\ -34}$~K, SO$_2$: $281\substack{+5 \\ -5}$~K, CH$_3$COCH$_3$: $377\substack{+26 \\ -25}$~K; see Tasa-Chaveli et al.\ in preparation).
The fitted velocity linewidths range from 8 to 11~km~s$^{-1}$, consistent with the NaCl and KCl detections by \citealp{Ginsburg2019, Ginsburg2023}. For CH$_3$CN, C$_2$H$_5$CN, CH$_3$OH and CH$_3$COCH$_3$, an additional narrower ($\Delta v\sim3$~km~s$^{-1}$) and colder ($T\sim150$~K) component, likely tracing extended gas surrounding the hot and compact disk, needs to be added.
Further results from the global fit of G351.77-mm1, together with the analysis of the chemical content for the other disks in the G351.77 region (see Fig.~\ref{fig:continuum}), will be presented in a forthcoming publication.

\begin{figure*}
    \centering
    \begin{subfigure}{0.24\textwidth}
        \centering
        \includegraphics[width=\textwidth]{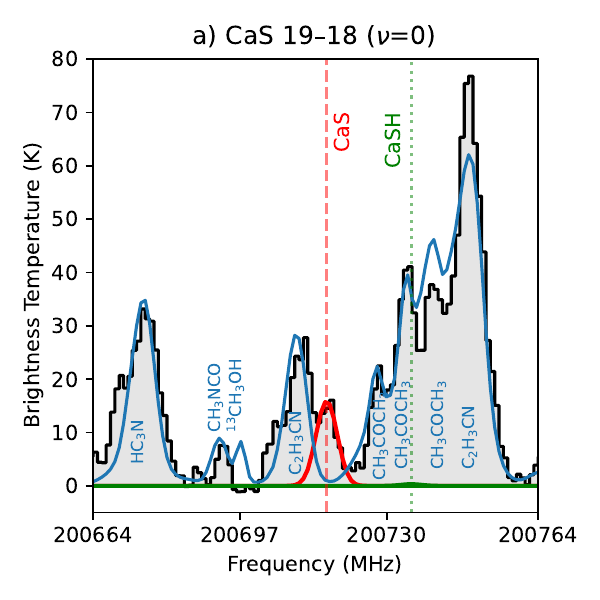} 
    \end{subfigure}  \hfill
    \begin{subfigure}{0.24\textwidth}
        \centering
        \includegraphics[width=\textwidth]{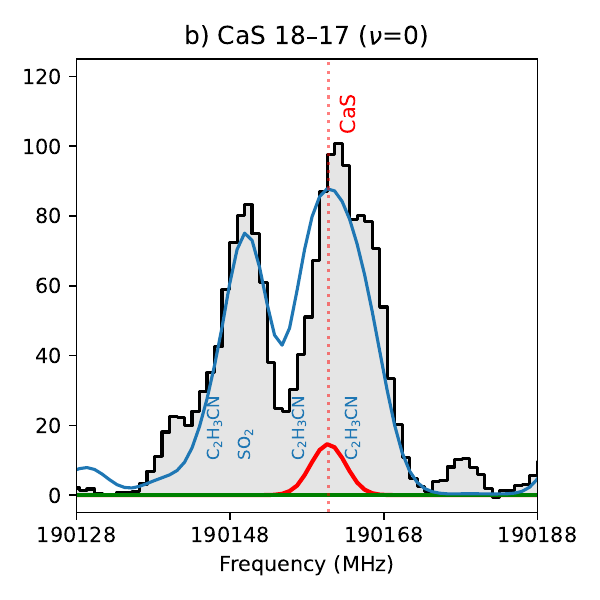}
    \end{subfigure}  \hfill
    \begin{subfigure}{0.24\textwidth}
         \centering
         \includegraphics[width=\textwidth]{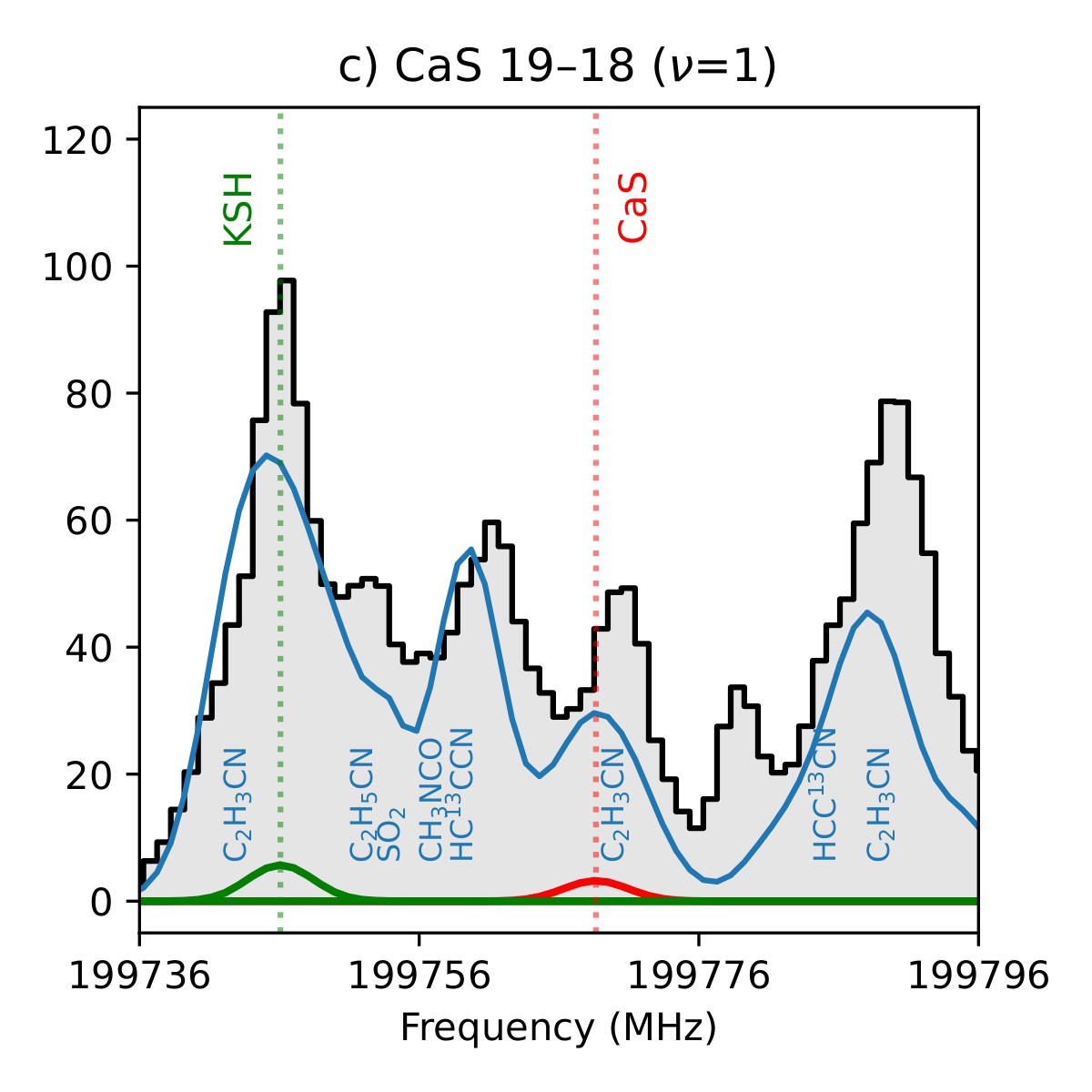} 
    \end{subfigure}
    \begin{subfigure}{0.24\textwidth}
        \centering
        \includegraphics[width=\textwidth]{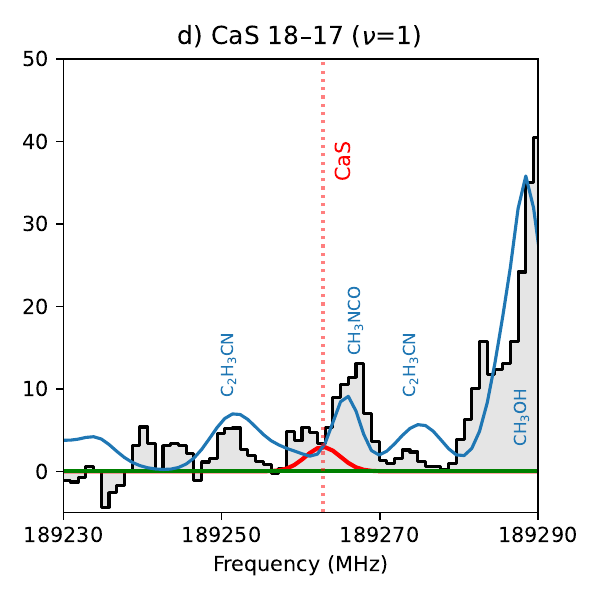}
    \end{subfigure}  
    
    \begin{subfigure}{0.24\textwidth}
        \centering
        \includegraphics[width=\textwidth]{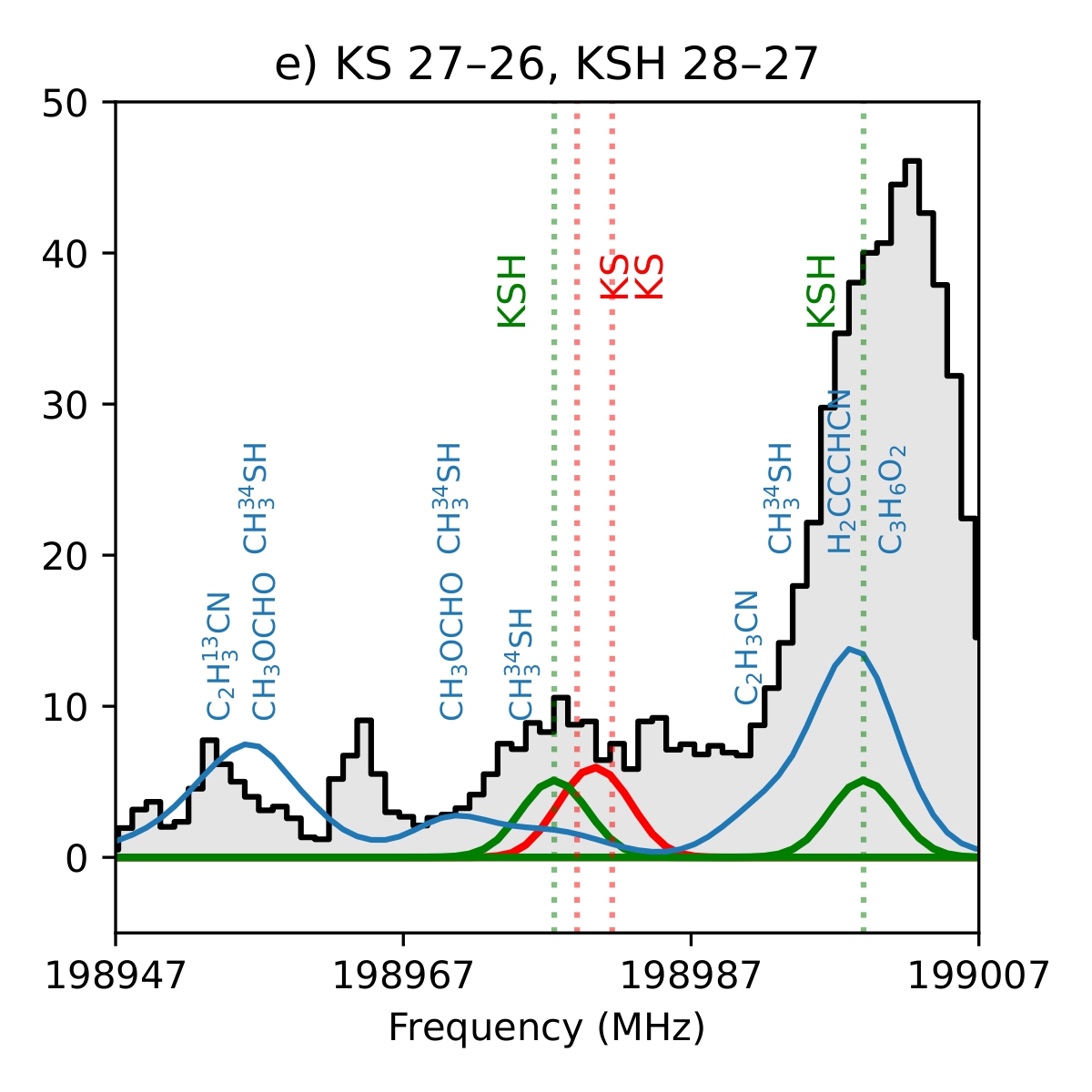} 
        
    \end{subfigure} 
    \begin{subfigure}{0.24\textwidth}
        \centering
        \includegraphics[width=\textwidth]{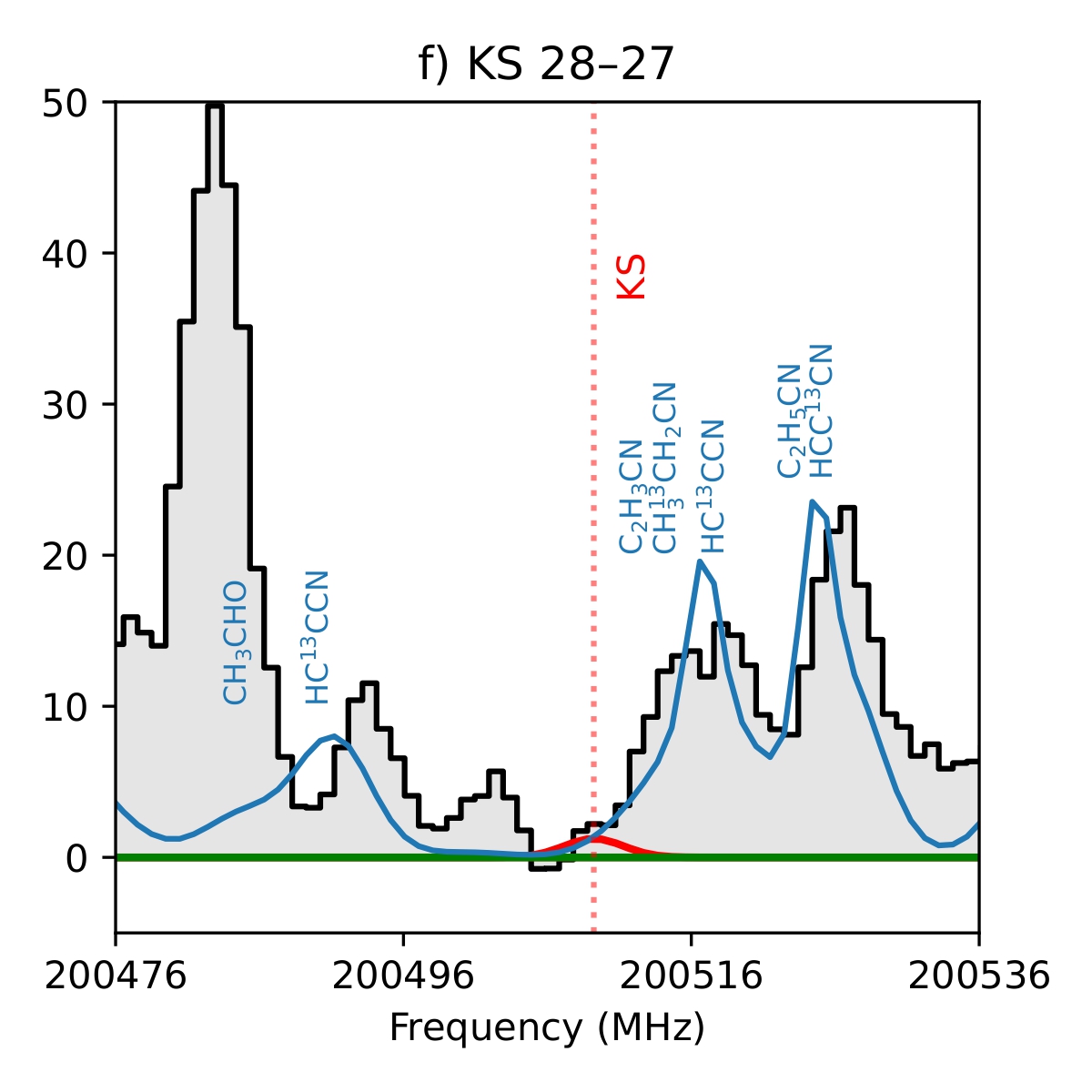}
    \end{subfigure}  
    \caption{Zoom in panels around the CaS and KS transitions included in the spectra of G351.77-mm1. The observed spectra are shown with grey-filled histograms. The synthetic spectrum generated without including the S-bearing refractory species is shown with a blue line, while the best fits of CaS and KS are shown in red. The best fits of the KSH and CaSH lines are shown with a green line (see Fig.~\ref{fig:zooms-CaSH} for additional frequency ranges covering non-detected CaSH transitions).}
    \label{fig:zooms}
\end{figure*}

%
\section{New species}\label{sec:new-species}
 
Once we identified the major components contributing to the observed spectrum, we used XCLASS to generate synthetic spectra of specific molecules and search for refractory species containing sulfur. The identifications of new species in crowded spectra is challenging and requires to establish some criteria \citep[see e.g.,][]{ReyMontejo2024}. Our criteria are as follows:
\begin{itemize}
\item[(i)] the frequencies we use are known with a precision better than $\pm$100~kHz, which is 10 times lower than the channel width (1.3~MHz);
\item[(ii)] the velocity offset and linewidth of the line candidates are in agreement with the values obtained for the species fitted in Sect.~\ref{sec:results};
\item[(iii)] we search in an interval of $\pm$7~MHz (similar to the typical line width observed in G351.77-mm1) to ensure that there is not any line of the species fitted in Sect.~\ref{sec:results}, nor their isotopologues, that can account for the emission;
\item[(iv)] we test that the emission cannot be accounted for by any other line of species detected in the interstellar and circumstelllar medium as listed in the CDMS catalogue, with transitions in the interval $\pm7$~MHz around the target spectral line feature;
\item[(v)] we also check for undetected species listed in catalogues that are expected to be abundant according to our current understanding of the chemistry;
\item[(vi)] we request for at least one clean (unblended) line in the observed band detected in our spectrum;
\item[(vii)] the generated spectra of the new species is consistent with the observations.
\end{itemize}
If all the conditions are fulfilled, we consider that our detection is reliable. If any of these conditions cannot be achieved, we consider that the detection is tentative. We are aware that since our spectral band is limited, there is always a margin for misidentification. Observations at different bands are required to claim a more robust detection. Following these criteria, we found convincing evidence for a reliable detection of CaS and tentative detections of KS and KSH, as explained below.
%
\subsection{Calcium sulfide, CaS}\label{sec:CaS}

Four transitions of CaS \citep[][]{Takano1989} are covered within the observed frequency range (two ground state and two vibrationally excited). These transitions (listed in Table~\ref{tab:info_my_lines} and highlighted in Fig.~\ref{fig:total-spectrum} with dashed lines) are shown in the top row of Fig.~\ref{fig:zooms}. We clearly identify (see panel~a of Fig.~\ref{fig:zooms}) one clean transition at 200.7141~GHz, CaS\,19$\rightarrow$18 ($\nu$=0), while the other three appear blended with transitions from other more abundant species (see panels~b, c and d).

In order to ensure that the line at 200.7141~GHz corresponds to the ground-state CaS\,19$\rightarrow$18 transition and to claim identification of CaS, we followed the criteria listed in Sect.~\ref{sec:new-species} and examined all the molecules at $\pm$7~MHz (see list in Table~\ref{tab:blending}). We first focused on species already detected in the ISM. Some of these potential contaminants are isotopologues of previously identified species. For instance, the isotopologues $^{13}$CH$_2$CHCN and $^{13}$CH$_3$CH$_2$CN were previously modeled and included in the synthetic spectrum shown in Fig.~\ref{fig:total-spectrum} proving that they do not contribute significantly to the emission line at 200.7141~GHz. In fact, our fitting allowed us to determine the isotopic ratios for these compounds. Other cases such as CH$_3^{34}$SH, C$_2$H$_5$C$^{15}$N and CH$_3$OD, not identified in the global spectrum fit, can not reproduce the observed spectral feature. If we try to fit the observed feature with any of these potential contaminants, other transitions expected to be in different parts of the spectra are not observed (see Fig.~\ref{fig:CaS200-blending} for details). This test is repeated considering different temperatures in the range 100 to 500~K to firmly exclude the presence of these contaminants. The same procedure was repeated for other species such as C$_3$H$_7$OH, C$_3$H$_7$CN, NH$_2$CH$_2$CH$_2$OH, CH$_3$COOH and CH$_3$CONH$_2$. These species are not detected down to the rms of the G351.77-mm1 spectrum. Finally, the remaining molecules with transitions within the $\pm7$~MHz frequency range are species that have never been detected in space, but are included in the molecular databases. We followed the same approach as indicated above, and determined that the observed spectral feature could not be attributed to any of these species. In particular, we have derived upper limits for the sulfur allotrope S$_4$ (see Table~\ref{tab:results}) using $T_{\rm rot}=400$~K since it is interesting for our discussion on the S-budget.

\begin{table*}
\begin{center}
\caption{Physical parameters of molecules (tentatively) detected toward G351.77-mm1 and discussed in this work\label{tab:results}}
\begin{tabular}{l c c c c c l}
\tableline\tableline \noalign{\smallskip}
  
& $N_\mathrm{mol}$
& $T_\mathrm{rot}$
& $\Delta v$
& $v_\mathrm{offset}$
& 
& 
\\
  Species
& (cm$^{-2}$)
& (K)
& (km~s$^{-1}$)
& (km~s$^{-1}$)
& Abundance
& Validation
\\
\tableline \noalign{\smallskip}
CaS      & \phs$4.7${\raisebox{0.5ex}{\tiny$\substack{+0.3 \\ -0.3}$}}$\times10^{14}$ & \phn400 & \phn8.0 & $-3.6$ & \phs$4.7${\raisebox{0.5ex}{\tiny$\substack{+0.3 \\ -0.3}$}}$\times10^{-11}$      & Reliable \\
CaSH     & $<$ $1.0\times10^{16}$ & \phn400 & \phn8.0 & $-3.6$ & $<$ $1.0\times10^{-9}$\phsup & Upper limit \\
KS       & \phs$1.9${\raisebox{0.5ex}{\tiny$\substack{+0.6\\ -0.4}$}}$\times10^{14}$ & \phn400 & \phn7.0 & $-3.6$ & \phs$1.9${\raisebox{0.5ex}{\tiny$\substack{+0.6\\ -0.4}$}}$\times10^{-11}$      & Tentative \\
KSH      & \phs$1.3${\raisebox{0.5ex}{\tiny$\substack{+0.4 \\ -0.4}$}}$\times10^{15}$ & \phn400 & \phn8.0 & $-3.6$ & \phs$1.3${\raisebox{0.5ex}{\tiny$\substack{+0.4 \\ -0.4}$}}$\times10^{-10}$      & Tentative \\
\tableline   
SO$_2$   & \phs$3.7${\raisebox{0.5ex}{\tiny$\substack{+0.1 \\ -0.7}$}}$\times10^{18}$ & \phn281${\raisebox{0.5ex}{\tiny$\substack{+46 \\ -48}$}}$ & \phn8.0 & $-3.6$ & \phs$3.7${\raisebox{0.5ex}{\tiny$\substack{+0.1\\ -0.7}$}}$\times10^{-7}$\phsup & Robust \\
CH$_3$SH & \phs$2.2${\raisebox{0.5ex}{\tiny$\substack{+2.5 \\ -0.8}$}}$\times10^{18}$ & \phn390${\raisebox{0.5ex}{\tiny$\substack{+210 \\ -80}$}}$ & 11.0    & $-3.6$ & \phs$2.2${\raisebox{0.5ex}{\tiny$\substack{+2.5 \\ -0.8}$}}$\times10^{-7}$\phsup & Robust \\
SiS      & \phs$1.9${\raisebox{0.5ex}{\tiny$\substack{+0.1 \\ -0.1}$}}$\times10^{17}$ & \phn400
 & 11.0    & $-3.6$ & \phs$1.9${\raisebox{0.5ex}{\tiny$\substack{+0.1 \\ -0.1}$}}$\times10^{-8}$\phsup & Robust \\
S$_4$    & $<$ $2.3\times10^{17}$ & \phn400 & \phn8.0 & $-3.6$ & $<$ $2.3\times10^{-8}$\phsup & Upper limit \\
\tableline   
\end{tabular}
\end{center}
\tablecomments{The species are labeled according to its validation (see last column) in different groups. {\it Robust}: undoubtful detection of a species commonly detected in star forming regions; {\it Reliable}: the detection is based on spectroscopic observations including at least one clean line; {\it Tentative}: all the lines in the band are blended or partially blended; {\it Upper limits}: All the lines are highly blended and only an upper limit can be derived.
Temperature and column density uncertainties were obtained through MCMC fits using XCLASS. In Table~\ref{tab:parameters-fitting} we list the derived column densities when different temperatures are assumed. This results in variations of about a factor two, slightly larger than the uncertainties derived from the MCMC analysis. The SiS species has been reported by \citet{Ginsburg2023}.}
\end{table*}

After this comprehensive analysis, we confirmed that CaS is the only molecule that can describe the spectral line observed at 200.7141~GHz. 
Figure~\ref{fig:zooms} shows in red the best fit that reproduces the observations, resulting in $N({\rm CaS})=4.7\substack{+0.3 \\ -0.3}\times10^{14}$~cm$^{-2}$ for $T_{\rm rot}=400~K$ (see Table~\ref{tab:results}). The column density uncertainty was obtained through a Markov chain Monte Carlo (MCMC) fit as implemented within XCLASS. 
We note that this uncertainty is slightly smaller than the range of column densities derived when a broad range of fixed temperatures is considered (see Table~\ref{tab:parameters-fitting}), suggesting that additional clean CaS transitions are needed to better constrain its temperature and column density. In Fig.~\ref{fig:total-spectrum}, we also show the residual spectra obtained after subtracting the synthetic spectra from the observed spectra. The residual spectra have an average intensity of $\approx$1.3~K in the frequency range $\pm50$~MHz around the CaS\,19$\rightarrow$18 transition. This is similar to the rms noise level of our observations (see Table~\ref{tab:spw-info}), and about ten times lower than the observed average brightness in the same frequency range ($\approx17$~K), confirming that the synthetic spectra reproduce the observed data with accuracy. It is also worth noting that the residuals are 15 times lower than the CaS\,19$\rightarrow$18 observed line intensity.

Taking into account that  only one intense unblended CaS line lies within our observed frequency band, we consider it as a ``reliable detection''. Observations of additional clean transitions are required for a more robust detection of this species. We have searched the ALMA archive for observations of the G351.77 region that could include additional CaS lines in their frequency coverage. We identified projects 2015.1.00496.S \citep[see][]{Beuther2019} and 2017.1.00237.S \citep[see][]{Ginsburg2023, Ishihara2024} to cover the CaS\,22$\rightarrow$21 transition at 232.372~GHz. It is worth noting that the angular resolution of these archival data is $\approx0\farcs05$, about ten times better than our observations; and the spectral resolution is $\approx5$~km~s$^{-1}$ and $\approx1.5$~km~s$^{-1}$, respectively, potentially limiting the identification and fitting of spectral lines in the first case. In Fig.~\ref{fig:zooms-otherdata} we show the observed spectrum extracted around the CaS\,22$\rightarrow$21 transition overlaid with the synthetic global (in blue) and CaS (in red) spectra generated using the best fit obtained from our observations (see Fig.~\ref{fig:total-spectrum} and Table~\ref{tab:results}). The archival spectra have been smoothed to a resolution of $0\farcs16$, comparable with our data (see Sect.~\ref{sec:observations}). The synthetic spectra are in agreement with the observed data, confirming the presence of the CaS\,22$\rightarrow$21 transition, although this line is partially blended with CH$_3$OCHO lines that show self-absorption features. In summary, the available archival data also support a reliable detection of CaS towards the G351.77-mm1 disk.

Finally, we have also searched for the hydrogenated form of CaS, i.e., CaSH \citep[][]{TalebBendiab1996}, but no unblended lines were found (see Fig.~\ref{fig:zooms-CaSH}). For this, we derived an upper limit $N({\rm CaSH})\leq10^{16}$~cm$^{-2}$, assuming $T_{\rm rot}=400$~K. 

%
\subsection{Potassium sulfide, KS}\label{sec:KS}

As shown in Fig.~\ref{fig:total-spectrum}, our observations cover three ground state transitions of KS \citep[see Table~\ref{tab:info_my_lines};][]{XinZiurys1998}, with two of them located very close at $\approx$198.97~GHz (see also Fig.~\ref{fig:zooms} panel e). Following the same approach as for CaS, we examined all molecules with transitions in the spectral vicinity $\pm$7~MHz (see Table~\ref{tab:blending}) to confirm that KS is the responsible species for the observed lines at $\approx$198.97~GHz. By generating synthetic spectra of these potential contaminants, we confirmed that none of them appears as a potential contributor to the observed lines. In the following, we discuss the only plausible candidates that could account for these features.

In particular, one interesting potential contributor is CH$_3^{34}$SH, since three of its most intense transitions fall within the $\pm$7~MHz frequency range around 198.97~GHz. A total of 24 lines of CH$_3^{34}$SH are present in the observed frequency range, although they appear highly blended with other species. We adopted the canonical isotopic ratio $^{32}$S/$^{34}$S=22.7 \citep{Lodders2019}, and generated a synthetic spectrum for CH$_3^{34}$SH based on the synthetic spectrum of CH$_3$SH that reproduces two lines in Fig.~\ref{fig:total-spectrum}. The CH$_3^{34}$SH lines contribute less than 20\% to the emission of the line feature detected at $\approx$198.97~GHz.

Among the species not yet detected in the ISM that could contribute to the observed lines, there is a KSH transition at 198.9751~GHz \citep[][]{Bucchino2013}. We have fitted the KS and KSH species together to find the set of parameters that best reproduce the observed lines. We present the best fit in the bottom row panels of Fig.~\ref{fig:zooms}. The fit results in $N(\rm KS)=1.9\substack{+0.6 \\ -0.4}\times10^{14}~cm^{-3}$ and $N(\rm KSH)=1.3\substack{+0.4 \\ -0.4}\times10^{15}~cm^{-3}$, with $T_{\rm rot}=400$~K (see Table~\ref{tab:parameters-fitting} for the fits obtained using different temperatures).
We also note that one l-SiC$_3$ \citep{Cernicharo2025} transition is blended with the KSH transition at 198.9751 GHz, which could also be responsible for this emission. In order to account for the observed emission at $\approx198.97$~GHz, we would need to assume $N({\rm l-SiC}_3)=2.8$--$4.1\times10^{18}$~cm$^{-2}$ for temperatures in the range 100 to 500~K.
This is more than one order of magnitude higher than $N({\rm SiS})=(1.9\pm 0.1)\times10^{17}$~cm$^{-2}$, which is not reasonable taking into account that SiS is one of the most abundant silicon-bearing species. Consequently, we conclude that the combination of KS and KSH is the most likely carrier of the observed feature. Due to the blending of these two species, we consider that both KS and KSH are ``tentatively detected'' in our data.
%
\section{Discussion}\label{sec:discussion}
Studying refractory molecules enables us to explore regions and conditions that were previously inaccessible. However, the use of these important chemical probes is limited by the small number of species that have been detected thus far. CaS together with SiS, MgS and NaS, and tentatively KS and KSH, are the only refractory species containing sulfur detected so far in the ISM. Moreover, gaseous species containing Ca have only been observed in the envelope of the carbon-rich evolved star IRC+10216 \citep{Cernicharo2019}. These authors detected calcium isocyanide (CaNC) and obtained upper limits for CaF, CaCl, CaC, CaCCH, and CaCH$_3$. \citet{ReyMontejo2024} presented the detection of the metal-bearing molecules NaS and MgS in the Galactic Center molecular cloud G+0.693$-$0.027 with abundances of a few $10^{-13}$. Up to our knowledge, these are the only detections of these species in star forming regions. In this work, we contribute to the search for sulfur-bearing refractory species with the first-time detection of CaS towards the disk G351.77-mm1. We derive a CaS abundance of $\sim10^{-11}$, for an H$_2$ column density\footnote{$N_{\rm H_2}$ was derived from the continuum map, using an optical depth $\kappa=0.01$~cm$^2$~g$^{-1}$ and a dust temperature $T_{\rm dust}=400$~K.} of $9.9\times10^{24}$~cm$^{-2}$. Although far lower than the abundances of SO$_2$ and CH$_3$SH, this CaS abundance in G351.77-mm1 is 2 orders of magnitude higher than the values found for the gaseous metallic species in the molecular cloud G+0.693$-$0.027.
The kinetic temperature of the molecular gas in G+0.693$-$0.027 ($\sim$70--150~K; see \citealt{Zeng2018}) and the grain sputtering produced by low-velocity shocks are expected to be the main desorption mechanisms. Contrary to that, higher kinetic temperatures are measured for the disk in G351.77-mm1, exceeding 1000~K \citep{Beuther2019}. At these high temperatures, thermal desorption and high-velocity shocks are expected to play a fundamental role in regulating the observed chemistry. Once the refractory atoms are released to the gas phase in these two regions (G351.77-mm1 and G+0.693$-$0.027), one expects that the subsequent reactions may follow different reaction paths, thus producing a differentiated chemistry between them. Interestingly, CaS is predicted to be one the most abundant refractory compounds in the inner envelope of evolved stars \citep{Agundez2020}. The detection of compounds containing SH radicals (CaSH and KSH) could provide additional insights into the origin of refractory gas-phase species since these compound could be formed on grain surfaces. The search and detection of new refractory species is thus necessary to disentangle the intricate chemical processes that can lead to their formation in different environments.

As stated earlier, identifying S-bearing refractory compounds could uncover a previously hidden sulfur reservoir that might address the `missing sulfur problem'. Based on this, we compare the column densities derived for CaS ($4.7\times10^{14}$~cm$^{-2}$) and KS ($<1.9\times10^{14}$~cm$^{-2}$) with those estimated for CH$_3$SH ($2.2\times10^{18}$~cm$^{-2}$) and SO$_2$ ($3.7\times10^{18}$~cm$^{-2}$), which are the most abundant sulfur carriers in our fitting. The contribution of refractory compounds to the sulfur budget appears to be negligible with column densities about 4 orders of magnitude lower than that of SO$_2$ and CH$_3$SH. We also note that the abundance of these two compounds are similar, and summed together, represent about 1/75 of the total sulfur budget; not enough to account for the missing sulfur abundance. 

%
\section{Summary and Conclusions}\label{sec:conclusions}

We present the first results of a high-angular resolution ALMA project aimed at searching for sulfur-bearing refractory species in star-forming regions. We present ALMA band 5 (186--203~GHz) observations of the disc around the massive young stellar object G351.77-mm1. We have identified an unblended transition of calcium sulfide (CaS) together with three additional partially blended transitions. The best fit results in an abundance relative to H$_2$ of $4.7\times10^{-11}$. Potassium sulfide (KS) and its hydrogenated form (KSH) have been tentatively detected with upper limits to their abundance of $<1.9\times10^{-11}$ and $<1.3\times10^{-10}$, respectively. Further observations in different bands covering additional transitions are required to establish robust detections of these species. The abundance of these sulfur-bearing refractory compounds is about 5 orders of magnitude lower than the elemental sulfur abundance, suggesting that these species do not constitute major reservoirs of sulfur at the spatial scales ($\approx300$~au) probed by our observations.

\begin{acknowledgments}

We thank the anonymous referee for the constructive comments that have helped to improve the content of the manuscript.
This project has received funding from the European Research Council (ERC) under the European Union’s Horizon Europe research and innovation programme ERC-AdG-2022 (GA No.\ 101096293). Funded by the European Union. Views and opinions expressed are however those of the author(s) only and do not necessarily reflect those of the European Union or the European Research Council Executive Agency. Neither the European Union nor the granting authority can be held responsible for them.
A.S.-M.\ acknowledges support from the RyC2021-032892-I grant funded by MCIN/AEI/10.13039/501100011033 and by the European Union `Next GenerationEU’/PRTR, as well as the program Unidad de Excelencia María de Maeztu CEX2020-001058-M, and support from the PID2023-146675NB-I00 (MCI-AEI-FEDER, UE).
A.F.\ and G.E.\ acknowledge funding support from project PID2022-137980NB-I00 funded by the Spanish Ministry of Science and Innovation/State Agency of Research MCIN/AEI/10.13039/501100011033 and by `ERDF A way of making Europe'.
The work at Universt\"at zu K\"oln was supported by the Deutsche Forschungsgemeinschaft (DFG) via the collaborative research center SFB 1601 (project ID 500700252) subprojects Inf (HSPM)
This paper makes use of the following ALMA data: ADS/JAO.ALMA\#2023.1.01382,  ADS/JAO.ALMA\#2015.1.00496 and
ADS/JAO.ALMA\\\#2017.1.00237. ALMA is a partnership of ESO (representing its member states), NSF (USA) and NINS (Japan), together with NRC (Canada), MOST and ASIAA (Taiwan), and KASI (Republic of Korea), in cooperation with the Republic of Chile. The Joint ALMA Observatory is operated by ESO, AUI/NRAO and NAOJ.
\end{acknowledgments}

\clearpage
\appendix
\restartappendixnumbering

%
\section{Supporting material: Tables and Figures}

\begin{figure*}
\centering
\includegraphics[width=1.0\textwidth]{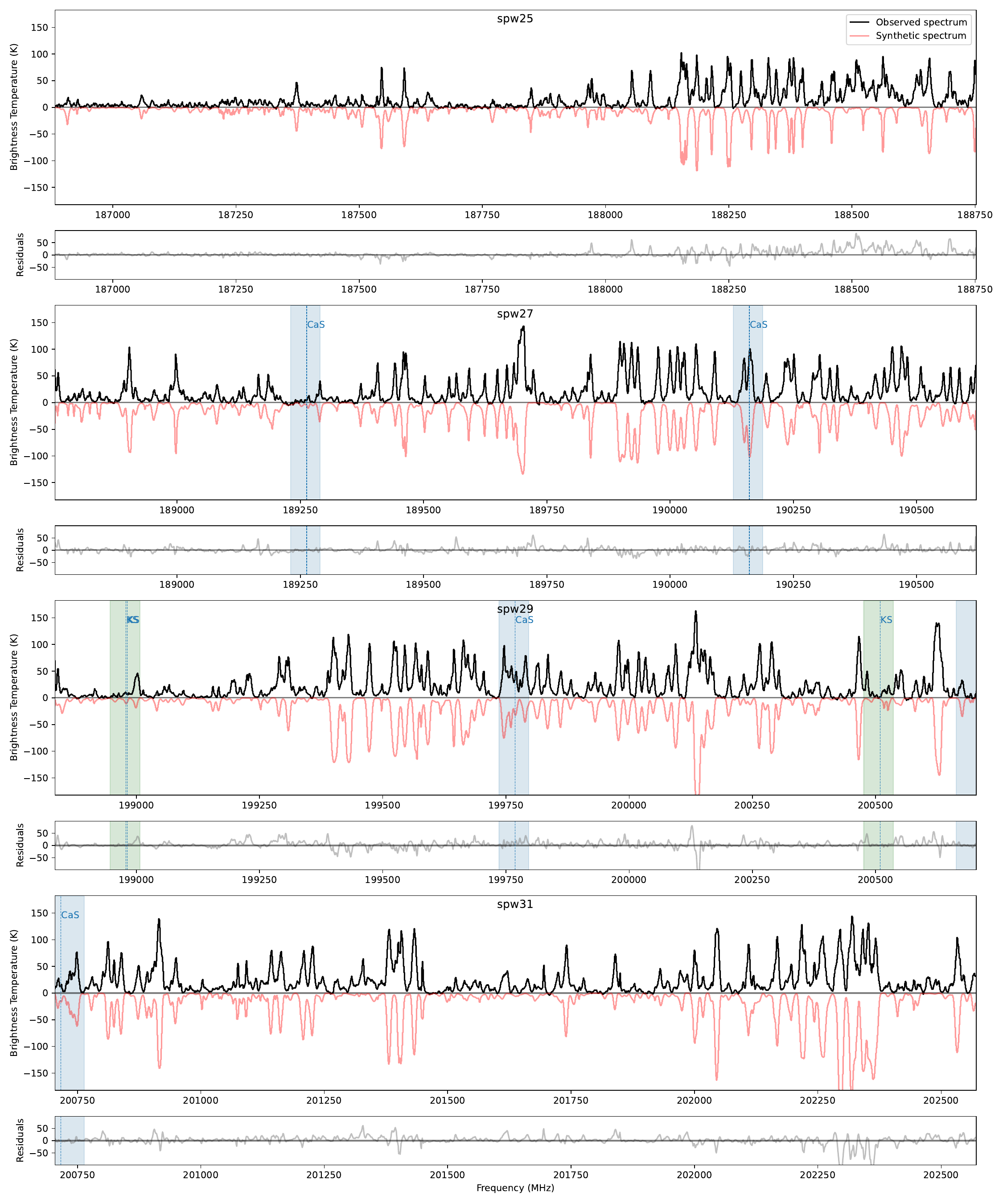}
\caption{Spectra towards G351.77-mm1, with the intensity in brightness temperature. ALMA observational data is shown in black. The synthetic spectrum including all identified molecular species (see Sect.~\ref{sec:results}) is shown in red, and with the intensity multiplied by a factor $-1$. Each major panel corresponds to a spectral window (see Table~\ref{tab:spw-info}), with the minor panels showing the residual spectra (observed minus synthetic). Frequencies of CaS and KS transitions are marked with vertical lines. Blue and green colored areas mark the zoomed-in frequency ranges around these transitions shown in Fig.~\ref{fig:zooms}.}
\label{fig:total-spectrum}
\end{figure*}

\begin{figure*}[t]
        \centering
 \includegraphics[width=0.49\textwidth]{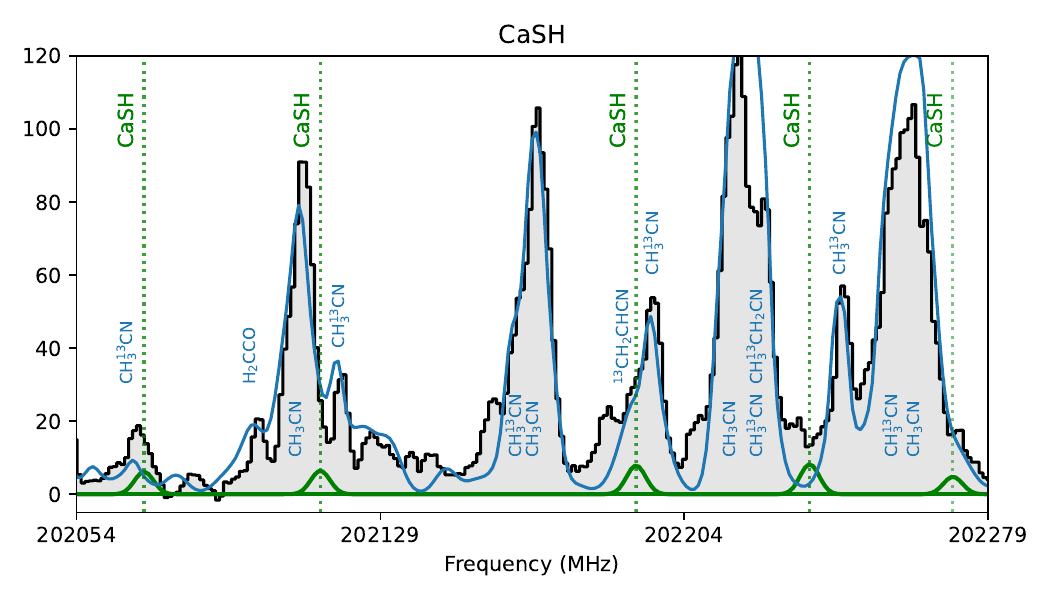}
 
    \caption{Zoom in panel around the CaSH transitions included in the spectra of G351.77-mm1. The observed spectrum is shown with grey-filled histograms. The synthetic spectrum generated without including the S-bearing refractory species is shown with a blue line, while the best fit of CaSH is shown in green.}
    \label{fig:zooms-CaSH}
\end{figure*}

\begin{figure*}[t]
\centering
\begin{subfigure}{0.49\textwidth}
\includegraphics[width=\textwidth]{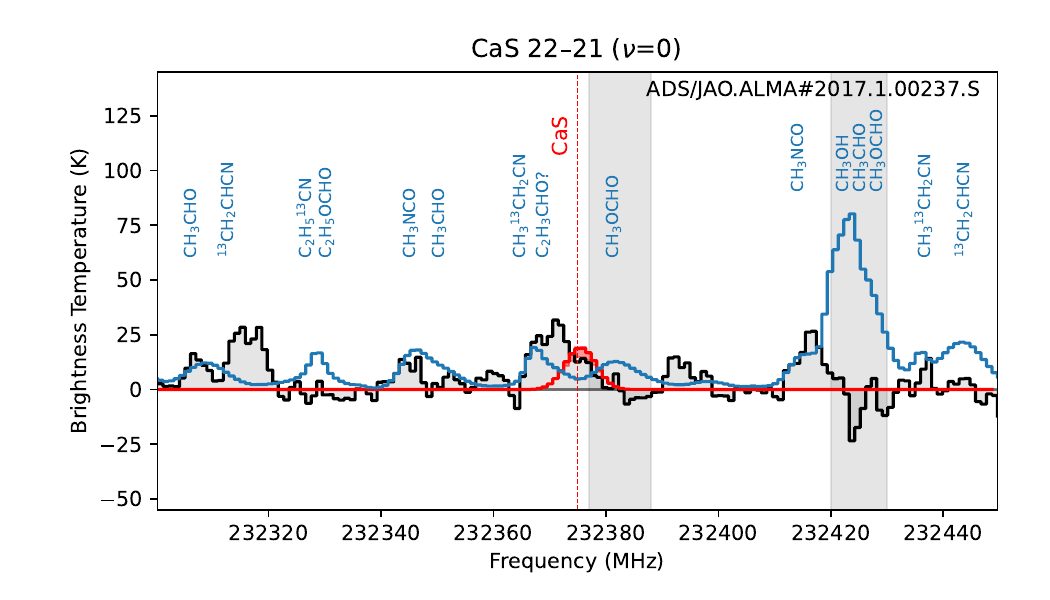} 
\end{subfigure}  
\hfill
\begin{subfigure}{0.49\textwidth}
\includegraphics[width=\textwidth]{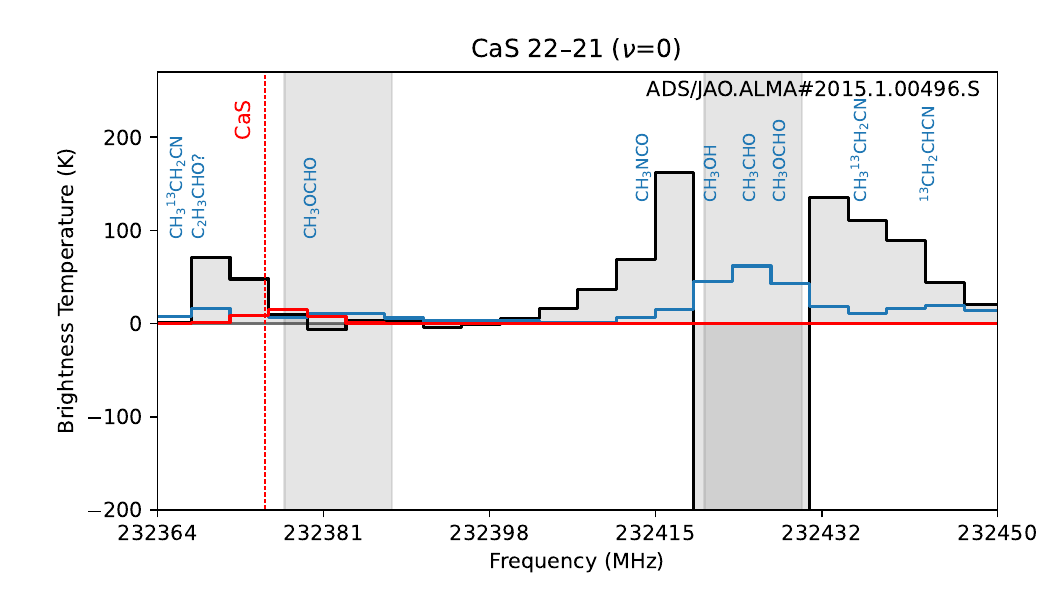}
\end{subfigure}
\caption{Spectra around the CaS\,22$\rightarrow$21 transition at 232.372~GHz observed towards G351.77-mm1, from ALMA archival data: (\textit{left panel}) project 2017.1.00237.S \citep[see][]{Ginsburg2023}, (\textit{right panel}) project 2015.1.00496.S \citep[see][]{Beuther2019}. The observed spectra are shown in black with grey-filled histograms. A Gaussian convolution was applied to match the beam size ($\sim0\farcs16$) of our observations, and the spectra was extracted towards the same position at $\alpha(\mathrm{J2000})=17^\mathrm{h}26^\mathrm{m}42.533^\mathrm{s}$ and $\delta(\mathrm{J2000})=-36^{\circ}9^{\prime}17.37^{\prime\prime}$.
The synthetic spectrum generated including all the identified molecules listed in Table~\ref{tab:other-species} is shown in blue, while the best CaS synthetic spectrum is shown in red. The red-shaded area around the red solid line represents the uncertainty derived from different fits assuming different rotational temperatures (see Table~\ref{tab:parameters-fitting}). The compound C$_2$H$_3$CHO is suggested as the carrier of the $\sim$232.365 GHz feature. However, this molecule does not exhibit intense lines in our observations, which precludes robust confirmation. The shaded vertical bands cover the molecular lines that appear self-absorbed.}
\label{fig:zooms-otherdata}
\end{figure*}

In the following, we present a series of additional tables and figures to complement those provided in the main sections of the publication. In more detail, Fig.~\ref{fig:total-spectrum} shows the continuum-subtracted spectrum extracted towards the position of the G351.77-mm1 disk (black line), and compares it with the synthetic spectrum (shown in the mirrored red spectrum). Table~\ref{tab:other-species} list the molecules that have been identified to reproduce the global spectrum shown in Fig.~\ref{fig:total-spectrum}. Figures~\ref{fig:zooms-CaSH} and \ref{fig:zooms-otherdata} complement Fig.~\ref{fig:zooms}. The first one presents a zoom in panel around the main CaSH transitions covered in our frequency setup (only upper limits are reported for this species), while the second shows the the synthetic spectrum generated for G351.77-mm1 overlaid on ALMA archival data covering the CaS\,22$\rightarrow$21 transition at 232.372~GHz.

Table~\ref{tab:info_my_lines} lists the transitions of the sulfur-bearing refractory species CaS, CaSH, KS and KSH studied in this work. The frequencies are obtained from the CDMS \citep[][]{Mueller2005} and JPL \citep[][]{Pickett1998} databases, which are also available through the Virtual Atomic and Molecular Data Center \citep[VAMDC;][]{Endres2016}. The laboratory spectroscopic work for these four species can be found in \citealp[][]{Takano1989, TalebBendiab1996, XinZiurys1998, Bucchino2013}.

Table~\ref{tab:blending} lists the potential contaminants to the main transitions of CaS and KS studied in this work. Complementing this table, Fig.~\ref{fig:CaS200-blending} shows that the potential contaminants listed in Table~\ref{tab:blending} must have a negligible contribution to the identified spectral line features to be consistent with the observed spectrum across the whole band.

Finally, in Table~\ref{tab:parameters-fitting} we list the fits obtained for the S-bearing species CaS, KS, CaSH, and KSH, after fixing the temperature to different values.

\begin{table*}[h!]
\begin{center}
\caption{List of molecules included in the XCLASS model of G351.77-mm1\label{tab:other-species}}
\begin{tabular}{c l}
\tableline\tableline \noalign{\smallskip}
  \# atoms
& Species (and isotopologues) included in the global fit shown in Fig.~\ref{fig:total-spectrum} \\
\tableline \noalign{\smallskip}
2   & NaCl (Na$^{37}$Cl) \\
3   & SO$_2$ ($^{34}$SO$_2$) \\
4   & D$_2$CS \\
5   & CH$_3$CN (CH$_3^{13}$CN) \\
    & CH$_3$OH ($^{13}$CH$_3$OH) \\
    & CH$_3$SH (CH$_3^{34}$SH) \\
    & HC$_3$N (H$^{13}$CCCN, HC$^{13}$CCN, HCC$^{13}$CN, H$^{13}$C$^{13}$CCN, H$^{13}$CC$^{13}$CN, HC$^{13}$C$^{13}$CN) \\
    & t-HCOOH \\
    & H$_2$CCO \\
6   & HC(O)NH$_2$ \\
    & H$_2$CCNH \\
7   & C$_2$H$_3$CN  ($^{13}$CH$_2$CHCN, CH$_2$$^{13}$CHCN, CH$_2$CH$^{13}$CN) \\
    & CH$_3$CHO \\
    & CH$_3$CCH  \\
    & CH$_3$NCO \\
8   & CH$_3$OCHO \\
    & H$_2$CCCHCN \\
9   & C$_2$H$_5$CN ($^{13}$CH$_3$CH$_2$CN, CH$_3$$^{13}$CH$_2$CN, CH$_3$CH$_2$$^{13}$CN) \\
10  & CH$_3$C(O)CH$_3$ \\
11  & C$_3$H$_6$O$_2$ (C$_2$H$_5$OCHO) \\
\tableline   
\end{tabular}
\end{center}
\end{table*}

\clearpage

\begin{table*}
\begin{center}
\footnotesize
\caption{Emission lines properties\label{tab:info_my_lines}}
\begin{tabular}{l l c c c c c}
\tableline\tableline \noalign{\smallskip}
  Species
& Transition $\nu$\,(J$\rightarrow$J')
& Frequency~(GHz)
& $\log_{10} A_\mathrm{ij}$~(s$^{-1}$)
& $E_\mathrm{u}$~(K)
& g$_\mathrm{u}$
\\
\tableline \noalign{\smallskip}
CaS  & $\nu$=1 J=18$\rightarrow$17                                                       & 189.26055 & -2.38 & 746.25  & 37 \\
     & $\nu$=0 J=18$\rightarrow$17                                                       & 190.15847 & -2.37 & 86.71   & 37 \\
     & $\nu$=1 J=19$\rightarrow$18                                                       & 199.76624 & -2.30 & 755.83  & 39 \\
     & $\nu$=0 J=19$\rightarrow$18                                                       & 200.71411 & -2.30 & 96.35   & 39 \\
CaSH & $\nu$=0 J=49/2$\rightarrow$47/2; N=24$\rightarrow$23; Ka=10; Kc=14$\rightarrow$13 & 200.73310  & -2.96 & 1482.2  & 50 \\
     & $\nu$=0 J=49/2$\rightarrow$47/2; N=24$\rightarrow$23; Ka=10; Kc=15$\rightarrow$14 & 200.73310  & -2.96 & 1482.2  & 50 \\
     & $\nu$=0 J=47/2$\rightarrow$45/2; N=24$\rightarrow$23; Ka=9; Kc=15$\rightarrow$14  & 200.98617 & -2.94 & 1225.57 & 48 \\
     & $\nu$=0 J=47/2$\rightarrow$45/2; N=24$\rightarrow$23; Ka=9; Kc=16$\rightarrow$15  & 200.98617 & -2.94 & 1225.57 & 48 \\
     & $\nu$=0 J=49/2$\rightarrow$47/2; N=24$\rightarrow$23; Ka=9; Kc=16$\rightarrow$15  & 201.03615 & -2.94 & 1225.61 & 50 \\
     & $\nu$=0 J=49/2$\rightarrow$47/2; N=24$\rightarrow$23; Ka=9; Kc=15$\rightarrow$14  & 201.03615 & -2.94 & 1225.61 & 50 \\
     & $\nu$=0 J=47/2$\rightarrow$45/2; N=24$\rightarrow$23; Ka=8; Kc=16$\rightarrow$15  & 201.26324 & -2.93 & 995.21  & 48 \\
     & $\nu$=0 J=47/2$\rightarrow$45/2; N=24$\rightarrow$23; Ka=8; Kc=17$\rightarrow$16  & 201.26324 & -2.93 & 995.21  & 48 \\
     & $\nu$=0 J=49/2$\rightarrow$47/2; N=24$\rightarrow$23; Ka=8; Kc=17$\rightarrow$16  & 201.31154 & -2.93 & 995.26  & 50 \\
     & $\nu$=0 J=49/2$\rightarrow$47/2; N=24$\rightarrow$23; Ka=8; Kc=16$\rightarrow$15  & 201.31154 & -2.93 & 995.26  & 50 \\
     & $\nu$=0 J=47/2$\rightarrow$45/2; N=24$\rightarrow$23; Ka=1; Kc=24$\rightarrow$23  & 201.49594 & -2.87 & 134.66  & 48 \\
     & $\nu$=0 J=47/2$\rightarrow$45/2; N=24$\rightarrow$23; Ka=7; Kc=17$\rightarrow$16  & 201.51108 & -2.91 & 791.35  & 48 \\
     & $\nu$=0 J=47/2$\rightarrow$45/2; N=24$\rightarrow$23; Ka=7; Kc=18$\rightarrow$17  & 201.51108 & -2.91 & 791.35  & 48 \\
     & $\nu$=0 J=49/2$\rightarrow$47/2; N=24$\rightarrow$23; Ka=1; Kc=24$\rightarrow$23  & 201.53607 & -2.87 & 134.71  & 50 \\
     & $\nu$=0 J=49/2$\rightarrow$47/2; N=24$\rightarrow$23; Ka=7; Kc=18$\rightarrow$17  & 201.55789 & -2.91 & 791.39  & 50 \\
     & $\nu$=0 J=49/2$\rightarrow$47/2; N=24$\rightarrow$23; Ka=7; Kc=17$\rightarrow$16  & 201.55789 & -2.91 & 791.39  & 50 \\
     & $\nu$=0 J=47/2$\rightarrow$45/2; N=24$\rightarrow$23; Ka=6; Kc=19$\rightarrow$18  & 201.72847 & -2.90 & 614.21  & 48 \\
     & $\nu$=0 J=47/2$\rightarrow$45/2; N=24$\rightarrow$23; Ka=6; Kc=18$\rightarrow$17  & 201.72847 & -2.90 & 614.21  & 48 \\
     & $\nu$=0 J=49/2$\rightarrow$47/2; N=24$\rightarrow$23; Ka=6; Kc=19$\rightarrow$18  & 201.77399 & -2.90 & 614.25  & 50 \\
     & $\nu$=0 J=49/2$\rightarrow$47/2; N=24$\rightarrow$23; Ka=6; Kc=18$\rightarrow$17  & 201.77399 & -2.90 & 614.25  & 50 \\
     & $\nu$=0 J=47/2$\rightarrow$45/2; N=24$\rightarrow$23; Ka=5; Kc=20$\rightarrow$19  & 201.91442 & -2.89 & 463.98  & 48 \\
     & $\nu$=0 J=47/2$\rightarrow$45/2; N=24$\rightarrow$23; Ka=5; Kc=19$\rightarrow$18  & 201.91442 & -2.89 & 463.98  & 48 \\
     & $\nu$=0 J=49/2$\rightarrow$47/2; N=24$\rightarrow$23; Ka=5; Kc=19$\rightarrow$18  & 201.95885 & -2.89 & 464.03  & 50 \\
     & $\nu$=0 J=49/2$\rightarrow$47/2; N=24$\rightarrow$23; Ka=5; Kc=20$\rightarrow$19  & 201.95885 & -2.89 & 464.03  & 50 \\
     & $\nu$=0 J=47/2$\rightarrow$45/2; N=24$\rightarrow$23; Ka=4; Kc=21$\rightarrow$20  & 202.06821 & -2.88 & 340.84  & 48 \\
     & $\nu$=0 J=47/2$\rightarrow$45/2; N=24$\rightarrow$23; Ka=4; Kc=20$\rightarrow$19  & 202.06821 & -2.88 & 340.84  & 48 \\
     & $\nu$=0 J=49/2$\rightarrow$47/2; N=24$\rightarrow$23; Ka=4; Kc=21$\rightarrow$20  & 202.11174 & -2.88 & 340.89  & 50 \\
     & $\nu$=0 J=49/2$\rightarrow$47/2; N=24$\rightarrow$23; Ka=4; Kc=20$\rightarrow$19  & 202.11174 & -2.88 & 340.89  & 50 \\
     & $\nu$=0 J=47/2$\rightarrow$45/2; N=24$\rightarrow$23; Ka=3; Kc=22$\rightarrow$21  & 202.18965 & -2.88 & 244.93  & 48 \\
     & $\nu$=0 J=47/2$\rightarrow$45/2; N=24$\rightarrow$23; Ka=3; Kc=21$\rightarrow$20  & 202.18973 & -2.88 & 244.93  & 48 \\
     & $\nu$=0 J=49/2$\rightarrow$47/2; N=24$\rightarrow$23; Ka=3; Kc=22$\rightarrow$21  & 202.23250  & -2.88 & 244.97  & 50 \\
     & $\nu$=0 J=49/2$\rightarrow$47/2; N=24$\rightarrow$23; Ka=3; Kc=21$\rightarrow$20  & 202.23258 & -2.88 & 244.97  & 50 \\
     & $\nu$=0 J=47/2$\rightarrow$45/2; N=24$\rightarrow$23; Ka=2; Kc=23$\rightarrow$22  & 202.26791 & -2.87 & 176.34  & 48 \\
     & $\nu$=0 J=47/2$\rightarrow$45/2; N=24$\rightarrow$23; Ka=2; Kc=22$\rightarrow$21  & 202.29526 & -2.87 & 176.34  & 48 \\
     & $\nu$=0 J=49/2$\rightarrow$47/2; N=24$\rightarrow$23; Ka=2; Kc=23$\rightarrow$22  & 202.31020  & -2.87 & 176.38  & 50 \\
     & $\nu$=0 J=47/2$\rightarrow$45/2; N=24$\rightarrow$23; Ka=0; Kc=24$\rightarrow$23  & 202.31292 & -2.87 & 121.41  & 48 \\
     & $\nu$=0 J=49/2$\rightarrow$47/2; N=24$\rightarrow$23; Ka=2; Kc=22$\rightarrow$21  & 202.33775 & -2.87 & 176.39  & 50 \\
     & $\nu$=0 J=49/2$\rightarrow$47/2; N=24$\rightarrow$23; Ka=0; Kc=24$\rightarrow$23  & 202.35466 & -2.87 & 121.46  & 50 \\
KS   & $\nu$=0 J=55/2$\rightarrow$53/2; N=27$\rightarrow$26                              & 198.97669 & -2.24 & 135.52  & 56 \\
     & $\nu$=0 J=55/2$\rightarrow$53/2; N=27$\rightarrow$26                              & 198.97913 & -2.24 & 135.52  & 56 \\
     & $\nu$=0 J=55/2$\rightarrow$53/2; N=28$\rightarrow$27                              & 200.50682 & -2.23 & 569.65  & 56 \\
KSH  & $\nu$=0 J=27$\rightarrow$26; Ka=10; Kc=17$\rightarrow$16                          & 189.54458 & -2.38 & 1527.7  & 55 \\
     & $\nu$=0 J=27$\rightarrow$26; Ka=10; Kc=18$\rightarrow$17                          & 189.54458 & -2.38 & 1527.7  & 55 \\
     & $\nu$=0 J=28$\rightarrow$27; Ka=3; Kc=26$\rightarrow$25                           & 198.85195 & -2.26 & 265.66  & 57 \\
     & $\nu$=0 J=28$\rightarrow$27; Ka=3; Kc=25$\rightarrow$24                           & 198.85200   & -2.26 & 265.66  & 57 \\
     & $\nu$=0 J=28$\rightarrow$27; Ka=2; Kc=27$\rightarrow$26                           & 198.97510  & -2.26 & 195.11  & 57 \\
     & $\nu$=0 J=28$\rightarrow$27; Ka=2; Kc=26$\rightarrow$25                           & 198.99660  & -2.26 & 195.12  & 57 \\
     & $\nu$=0 J=28$\rightarrow$27; Ka=0; Kc=28$\rightarrow$27                           & 199.06132 & -2.26 & 138.62  & 57 \\
     & $\nu$=0 J=28$\rightarrow$27; Ka=1; Kc=27$\rightarrow$26                           & 199.74368 & -2.25 & 153.24  & 57 \\ \hline
\tableline   
\end{tabular}
\end{center}
\end{table*}

\begin{table*}
\normalsize
\begin{center}
\caption{List of potential contaminants (within $\pm$7~MHz) of the main CaS and KS transitions\label{tab:blending}}
\begin{tabular}{l l l}
\tableline\tableline \noalign{\smallskip}
  Line
& Potential contaminants
& Comments
\\
\tableline \noalign{\smallskip}
CaS\,(19--18)                 & $^{13}$CH$_2$CHCN             & Included in the global fit in Fig.~\ref{fig:total-spectrum} \\
\phn\phn at 200.7141~GHz      & $^{13}$CH$_3$CH$_2$CN         & Included in the global fit in Fig.~\ref{fig:total-spectrum} \\
                              & $^{13}$CH$_3$$^{13}$CH$_2$CN  & Included in the global fit in Fig.~\ref{fig:total-spectrum} \\
                              & $^{13}$CH$_3$CH$_2$$^{13}$CN  & Included in the global fit in Fig.~\ref{fig:total-spectrum} \\
                              & CH$_3$$^{13}$CH$_2$$^{13}$CN  & Included in the global fit in Fig.~\ref{fig:total-spectrum} \\
                              & C$_2$H$_5$OCHO               & Included in the global fit in Fig.~\ref{fig:total-spectrum} \\
                              & C$_2$H$_5$C$^{15}$N           & Negligible / non-detected (see Fig.~\ref{fig:CaS200-blending}) \\
                              & CH$_3^{34}$SH                 & Negligible / non-detected (see Fig.~\ref{fig:CaS200-blending}) \\
                              & CH$_3$OD                      & Negligible / non-detected (see Fig.~\ref{fig:CaS200-blending}) \\
                              & CHD$_2$OH                     & Negligible / non-detected (see Fig.~\ref{fig:CaS200-blending}) \\
                              & CH$_3$COOH                    & Negligible / non-detected (see Fig.~\ref{fig:CaS200-blending}) \\
                              & CH$_3$CONH$_2$                & Negligible / non-detected (see Fig.~\ref{fig:CaS200-blending}) \\
                              & $^{13}$CH$_2$(OH)CHO          & Negligible / non-detected (see Fig.~\ref{fig:CaS200-blending}) \\
                              & C$_3$H$_7$OH                  & Negligible / non-detected (see Fig.~\ref{fig:CaS200-blending}) \\
                              & C$_3$H$_7$CN                  & Negligible / non-detected (see Fig.~\ref{fig:CaS200-blending}) \\
                              & NH$_2$CH$_2$CH$_2$OH          & Negligible / non-detected (see Fig.~\ref{fig:CaS200-blending}) \\
\tableline
KS\,(27-26)                   & $^{13}$CH$_2$CHCN             & Included in the global fit in Fig.~\ref{fig:total-spectrum} \\
\phn\phn at 198.9767~GHz      & KSH                           & Fitted together with KS in Fig.~\ref{fig:zooms} \\
\phn\phn and at 198.9791~GHz  & C$_2$H$_5$C$^{15}$N           & Not detected \\
                              & CH$_3^{34}$SH                 & Not detected \\
                              & CH$_3$SD                      & Not detected \\
                              & CH$_3$O$^{13}$CHO             & Not detected \\
                              & t-H$^{13}$COOH                & Not detected \\
                              & CH$_3$COOH                    & Not detected \\
                              & CH$_3$CONH$_2$                & Not detected \\
                              & (CH$_2$OH)$_2$                & Not detected \\
                              & CH$_2$OHCHO                   & Not detected \\
                              & C$_2$H$_5$OH                  & Not detected \\
                              & CH$_3$CH$_2$OD                & Not detected \\
                              & C$_2$H$_5$NCO                 & Not detected \\
                              & c-C$_3$H$_5$CN                & Not detected \\
                              & C$_3$H$_7$OH                  & Not detected \\
                              & C$_3$H$_7$CN                  & Not detected \\
                              & C$_6$H$_5$CN                  & Not detected \\
                              & H$^{18}$ONO$_2$               & Not detected \\
                              & HC(S)CN                       & Not detected \\
                              & Z-HC$_2$CHCHCN                & Not detected \\
                              & H$_2$C(CN)$_2$                & Not detected \\
                              & H$_2$NCH$_2$COOH              & Not detected \\
                              & H$_2$N$^{13}$CH$_2$CN         & Not detected \\
                              & D$_2$NCH$_2$CN                & Not detected \\
                              & NH$_2$CH$_2$CH$_2$OH          & Not detected \\
                              & l-SiC$_3$         
                              & Not detected \\                              
\tableline   
\end{tabular}
\end{center}
\end{table*}
\clearpage
\begin{table*}
\begin{center}
\caption{Best-fit column densities derived for different fixed rotational temperatures\label{tab:parameters-fitting}}
\begin{tabular}{l c c c c}
\tableline\tableline \noalign{\smallskip}
& $T_\mathrm{rot}$
& $\Delta v$
& $v_\mathrm{offset}$
& $N_\mathrm{mol}$
\\
  Species
& (K)
& (km~s$^{-1}$)
& (km~s$^{-1}$)
& (cm$^{-2}$)
\\
\tableline \noalign{\smallskip}
CaS      & 100 & \phn8.0 & $-3.6$ & $2.5\times10^{14}$ \\
                        &  300    &  &  & $3.5\times10^{14}$  \\
                        &  400    &  &  & $4.7\times10^{14}$  \\
                        &  500    &  &  & $6.0\times10^{14}$  \\ 
                        &  700    &  &  & $9.4\times10^{14}$  \\\tableline \noalign{\smallskip} 
KS      & 100 & \phn7.0 & $-3.6$& $1.1\times10^{14}$\\
                        &  300    &  &  & $1.6\times10^{14}$  \\
                        &  400    &  &  & $2.0\times10^{14}$  \\
                        &  500    &  &  & $2.4\times10^{14}$  \\ 
                        &  700    &  &  & $3.2\times10^{14}$ \\\tableline \noalign{\smallskip}
KSH      & 100 & \phn8.0 & $-3.6$ & $7.2\times10^{14}$   \\
                        &  300    &  &  & $9.8\times10^{14}$  \\
                        &  400    &  &  & $1.3\times10^{15}$  \\
                        &  500    &  &  & $1.7\times10^{15}$  \\
                        &  700    &  &  & $2.5\times10^{15}$  \\ \tableline
SiS      & 100 & \phn11.0 & $-3.6$ & $2.1\times10^{17}$   \\
                        &  300    &  &  & $1.5\times10^{17}$  \\
                        &  400    &  &  & $1.9\times10^{17}$  \\
                        &  500    &  &  & $2.3\times10^{17}$  \\
                        &  700    &  &  & $3.3\times10^{17}$  \\
\tableline
\end{tabular}
\end{center}
\end{table*}
\clearpage
\begin{figure*}
    \centering
    \begin{subfigure}{\textwidth}
    \includegraphics[width=0.9\linewidth]{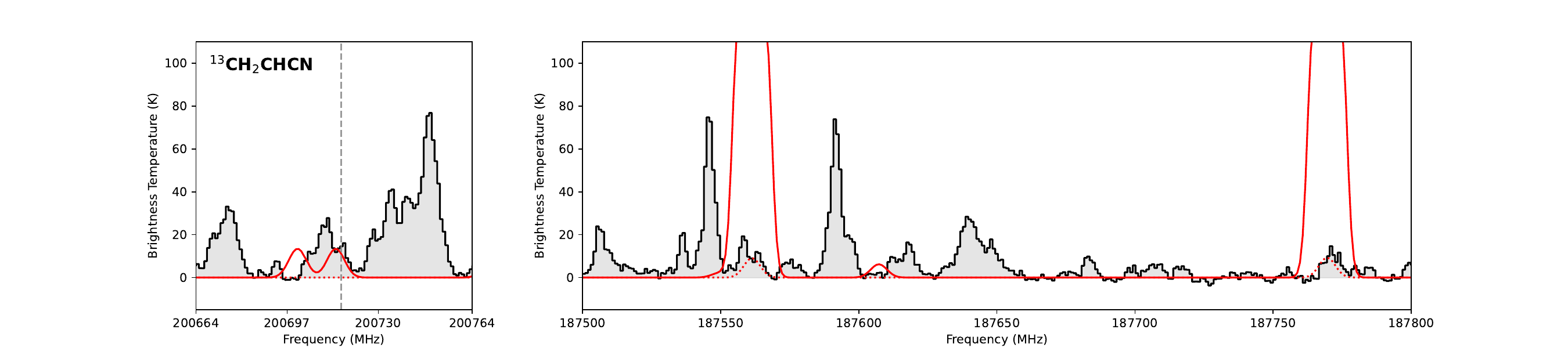}  
    \end{subfigure}
    \begin{subfigure}{\textwidth}
    \includegraphics[width=0.9\linewidth]{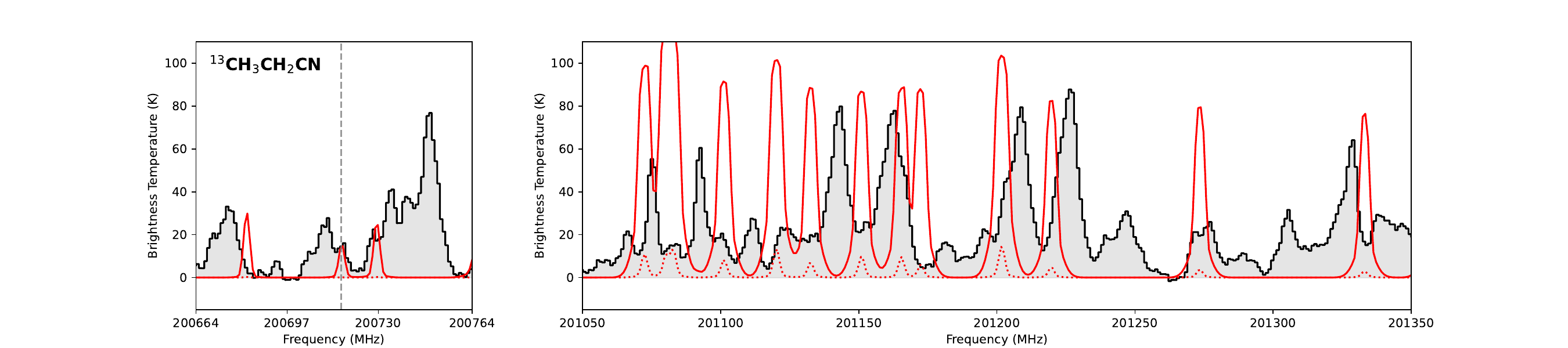}  
    \end{subfigure}    
    \begin{subfigure}{\textwidth}
    \includegraphics[width=0.9\linewidth]{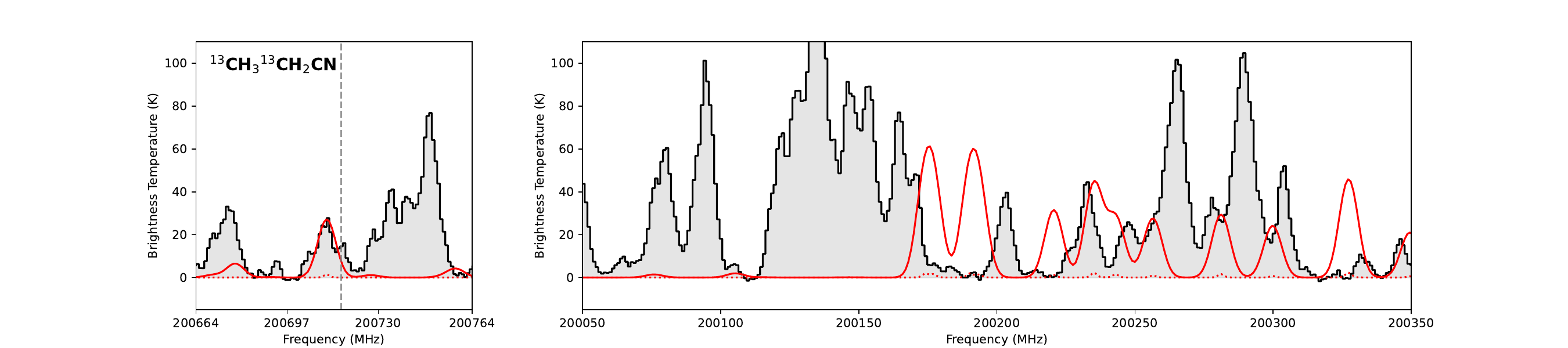}  
    \end{subfigure}
    
    \begin{subfigure}{\textwidth}
    \includegraphics[width=0.9\linewidth]{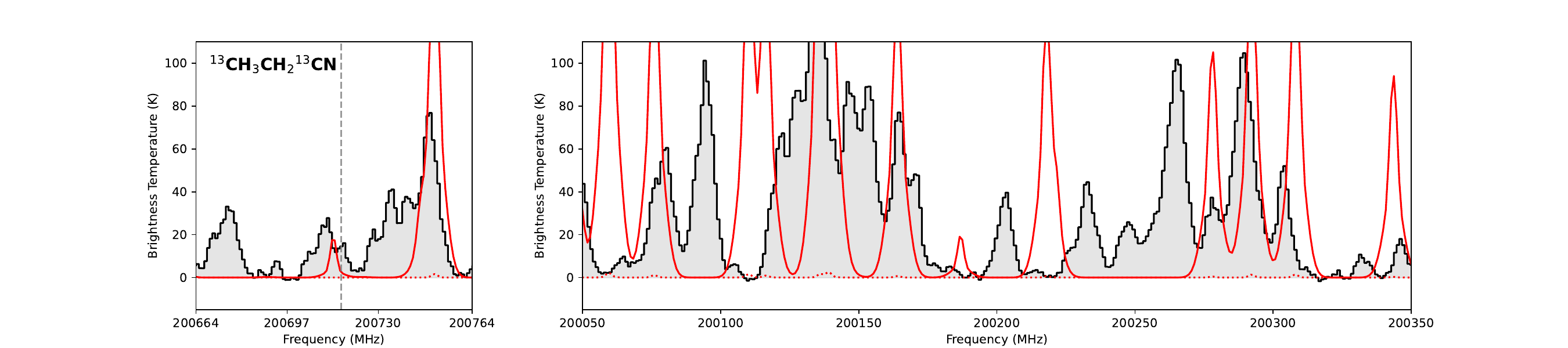}  
    \end{subfigure}
    
    \begin{subfigure}{\textwidth}
    \includegraphics[width=0.9\linewidth]{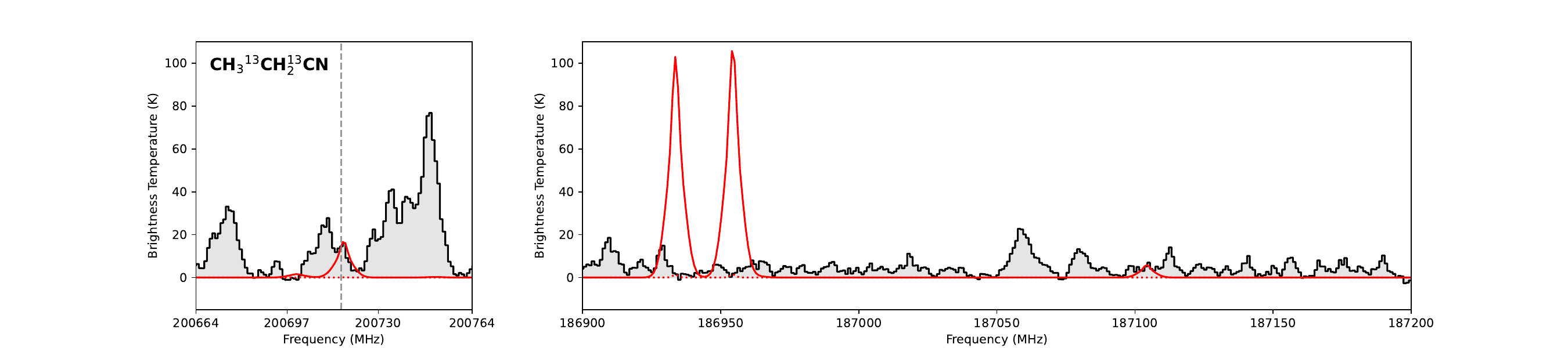}  
    \end{subfigure}
\caption{Synthetic spectra of potential contaminants of CaS\,(19--18; $\nu$=0) listed in Table~\ref{tab:blending}. The left panel in each row shows the observed spectrum (grey histogram) around the frequency of the CaS\,(19--18; $\nu$=0) transition, which is highlighted with a vertical dashed line. The red solid line correspond to the maximum intensity of each potential contaminant (labeled in black in the top-left corner of each panel) required to fully account for the observed line feature. These species are discarded on the basis of the predicted emission in other parts of the spectra (see right panels) which would be too bright to reproduce the observations. Red, dotted lines represent the maximum intensity of the potential contaminants, allowed by the upper limits derived from transitions distributed throughout the observed spectral range.}
\label{fig:CaS200-blending}
\end{figure*}
%
\begin{figure*}
\ContinuedFloat 
    \centering

    \begin{subfigure}{\textwidth}
    \includegraphics[width=0.9\linewidth]{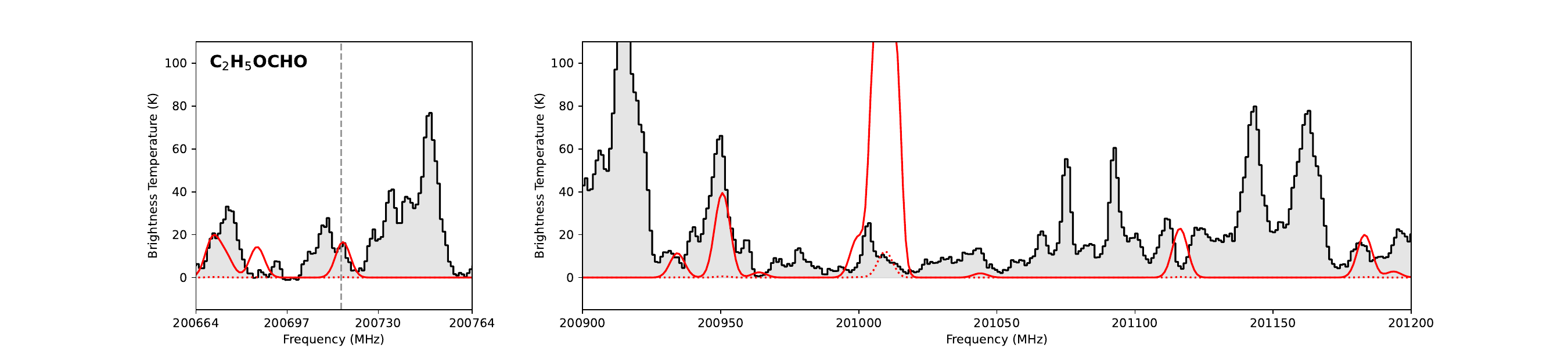}  
    \end{subfigure}
    
    \begin{subfigure}{\textwidth}
    \includegraphics[width=0.9\linewidth]{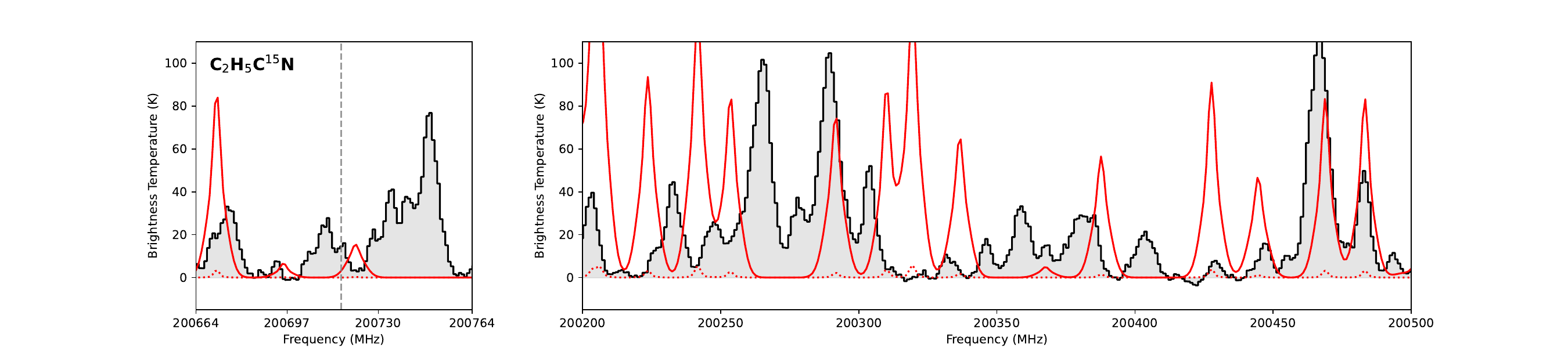}  
    \end{subfigure}
    
    \begin{subfigure}{\textwidth}
    \includegraphics[width=0.9\linewidth]{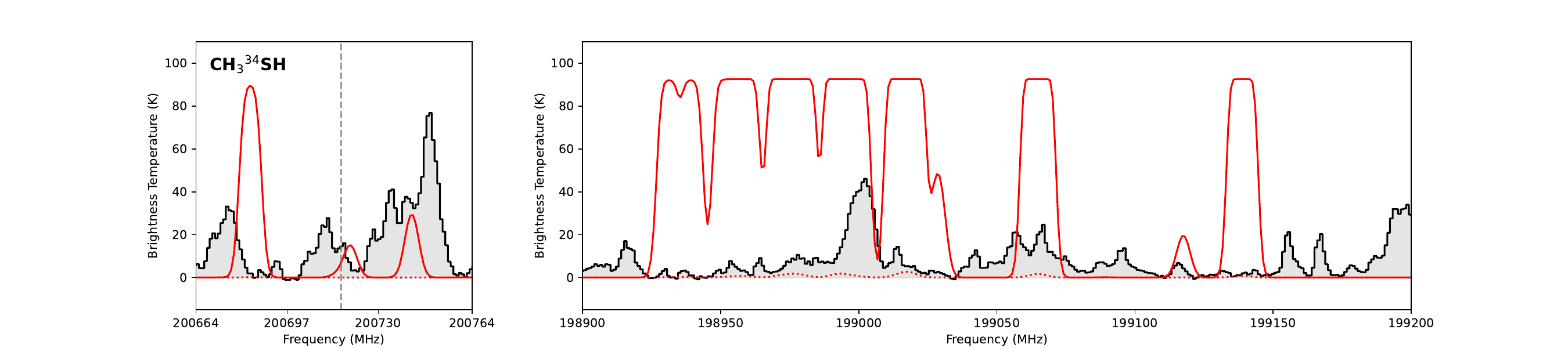}  
    \end{subfigure}

    \begin{subfigure}{\textwidth}
    \includegraphics[width=0.9\linewidth]{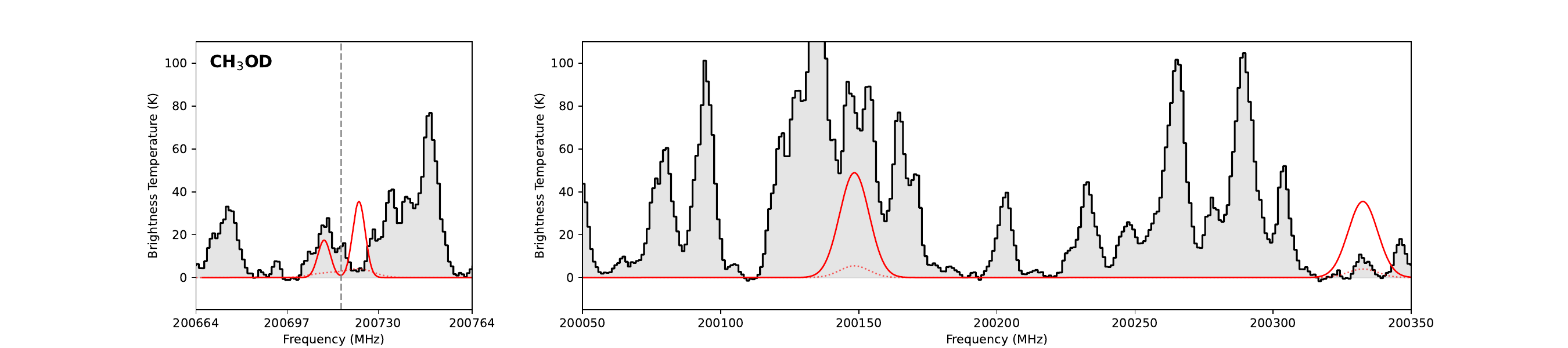}  
    \end{subfigure}
    
    \begin{subfigure}{\textwidth}
    \includegraphics[width=0.9\linewidth]{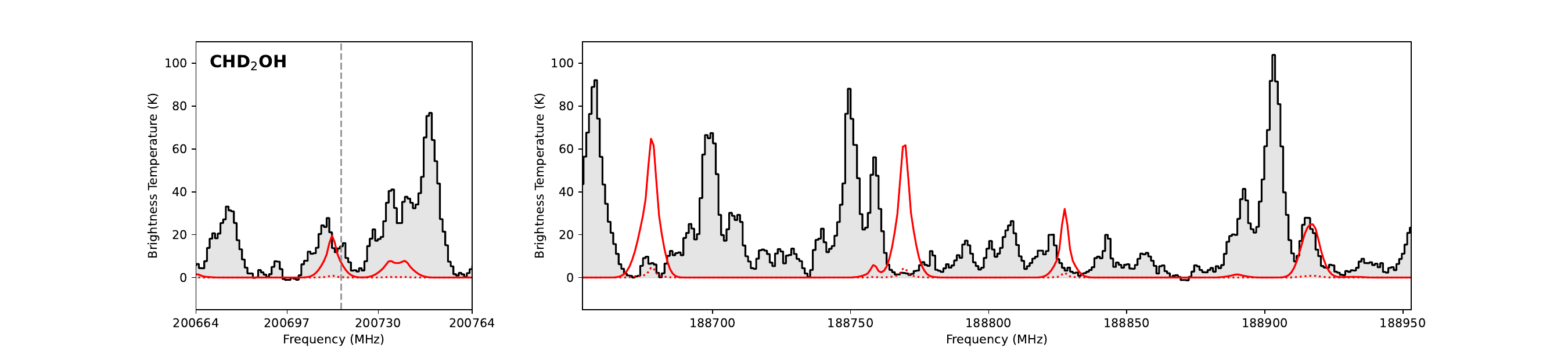}  
    \end{subfigure}
    
    \begin{subfigure}{\textwidth}
    \includegraphics[width=0.9\linewidth]{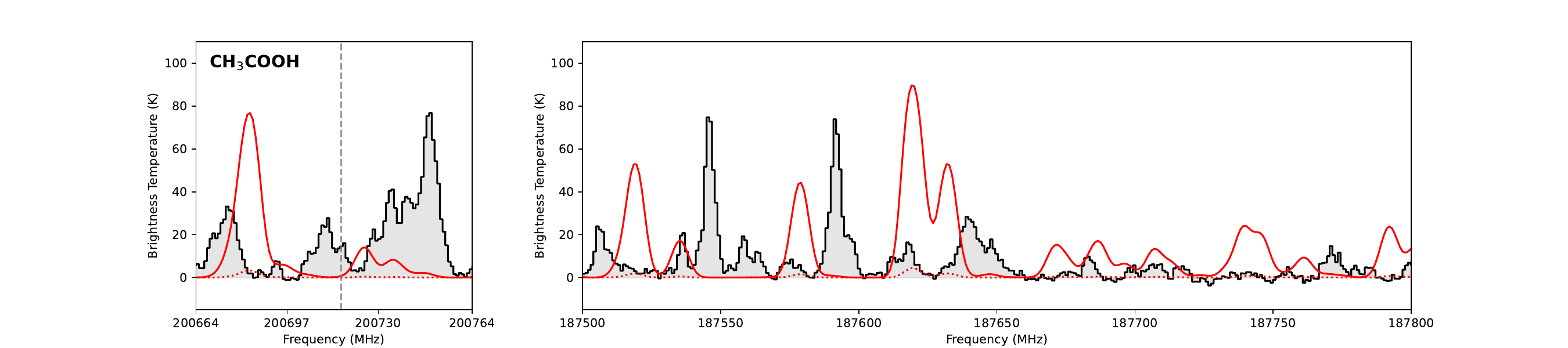}  
    \end{subfigure} 
\caption{\textit{Continued.}}
\end{figure*}
%
\begin{figure*}
\ContinuedFloat 
    \centering   
    \begin{subfigure}{0.9\textwidth}
    \includegraphics[width=\linewidth]{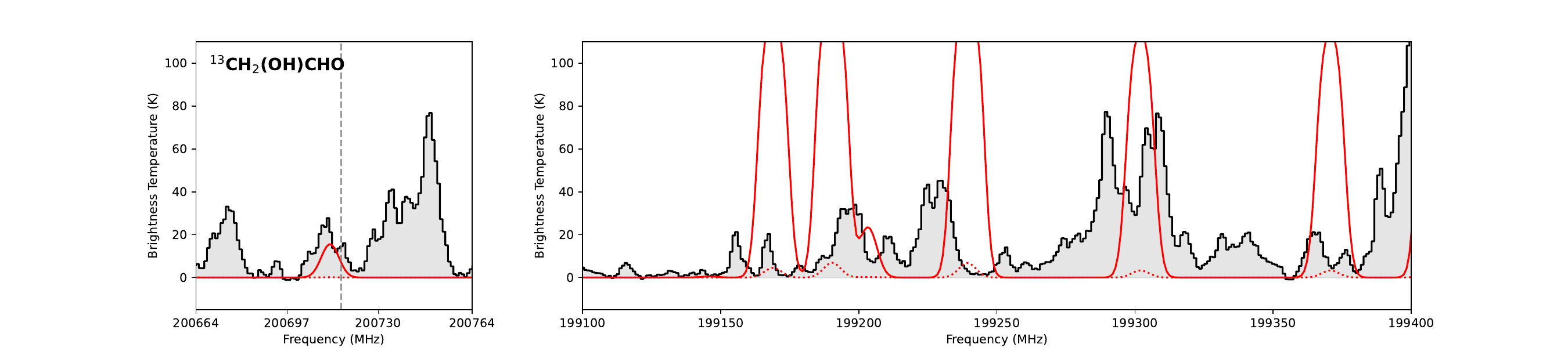}  
    \end{subfigure} 

    \begin{subfigure}{0.9\textwidth}
    \includegraphics[width=\linewidth]{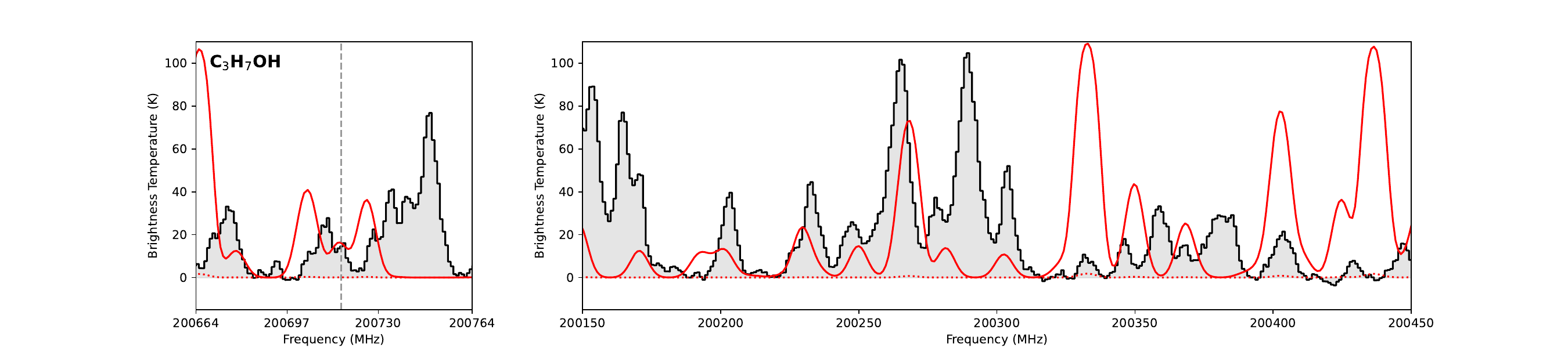}  
    \end{subfigure} 
    
    \begin{subfigure}{0.9\textwidth}
    \includegraphics[width=\linewidth]{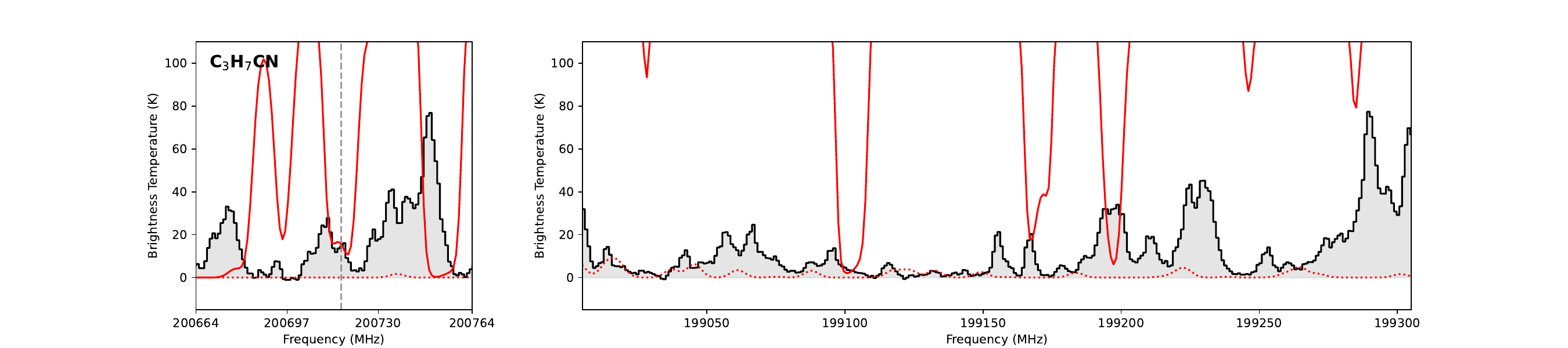}  
    \end{subfigure} 

    \begin{subfigure}{0.9\textwidth}
    \includegraphics[width=\linewidth]{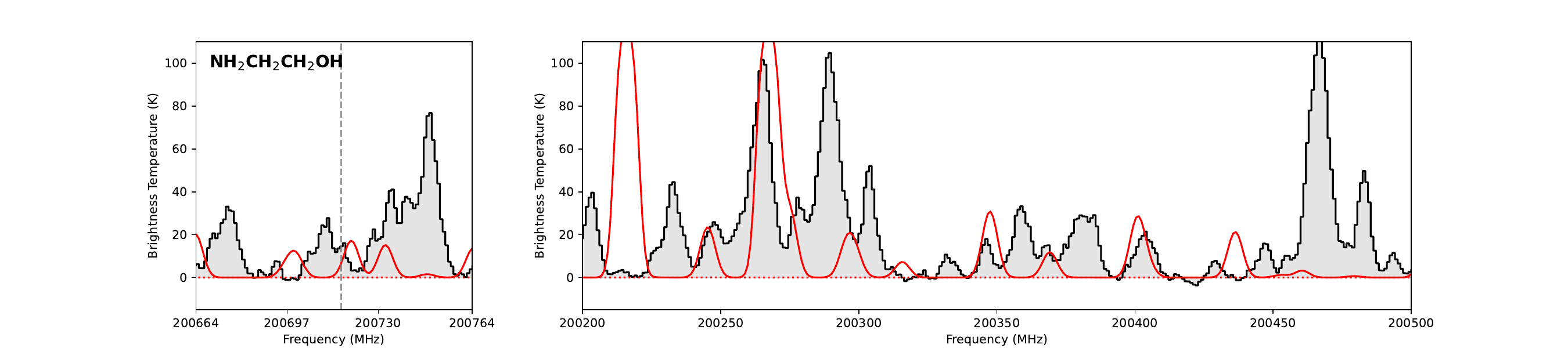}  
    \end{subfigure} 
\caption{\textit{Continued.}}
\end{figure*}

\clearpage
\bibliography{refs}{}
\bibliographystyle{aasjournalv7}

\end{document}